\documentclass[12pt,preprint]{aastex}
\usepackage{color}

\newcommand{\beq}{\begin{equation}}
\newcommand{\eeq}{\end{equation}}
\newcommand{\bdm}{\begin{displaymath}}
\newcommand{\edm}{\end{displaymath}}
\newcommand{\bea}{\begin{eqnarray}}
\newcommand{\eea}{\end{eqnarray}}
\newcommand{\bt}{\begin{tabular}}
\newcommand{\et}{\end{tabular}}
\newcommand{\kv}{{\bf k}}

\newcommand{\tr}{\tilde{r}}

\newcommand{\smin}{\mbox{\tiny{min}}}

\def\d{\delta}
\def\Ms{\, h^{-1} \, {\rm M}_{\odot}}
\def\Mpc{\, h^{-1} \, {\rm Mpc}}

\def\icMpc{\, h^3 \, {\rm Mpc}^{-3}}

\def\cGpc{\, h^{-3} \, {\rm Gpc}^3}

\def\Ng{N_{200}}

\begin{document}

\title{The Correlation Function of Optically Selected Galaxy Clusters\\
 in the Sloan Digital Sky Survey}

\author{Juan Estrada}
\email{estrada@fnal.gov}
\affil{Center for Particle Astrophysics, Fermi National Accelerator 
Laboratory, Batavia, IL 60510-0500}
\author{Emiliano Sefusatti}
\affil{Center for Particle Astrophysics, Fermi National Accelerator 
Laboratory, Batavia, IL 60510-0500}
\author{Joshua A. Frieman}
\affil{Center for Particle Astrophysics, Fermi National Accelerator 
Laboratory, Batavia, IL 60510-0500}
\affil{Kavli Institute for Cosmological Physics and Department of 
Astronomy \& Astrophysics, The University of Chicago, Chicago,IL 60637}

\begin{abstract}

We measure the two-point spatial correlation function for clusters selected from the photometric MaxBCG galaxy cluster catalog for the Sloan Digital Sky Survey (SDSS). We evaluate the correlation function for several cluster samples using different cuts in cluster richness. Fitting the results to power-laws, $\xi_{cc}(r) = (r/R_0)^{-\gamma}$, the estimated correlation length $R_0$ as a function of richness is broadly consistent with previous cluster observations and with expectations from N-body simulations. We study how the linear bias parameter scales with richness and compare our results to theoretical predictions. Since these measurements extend to very large scales, we also compare them to models that include the baryon acoustic oscillation feature and that account for the smoothing effects induced by errors in the cluster photometric redshift estimates. For the largest cluster sample, corresponding to a richness threshold of $\Ng\ge 10$, we find only weak evidence, of about $1.4-1.7\sigma$ significance, for the baryonic acoustic oscillation signature in the cluster correlation function.

\end{abstract}


\section{Introduction}

Galaxy clusters have long been recognized as powerful cosmological probes \citep{cluster_cosmo,RozoEtal2007cosmo}. In particular, measurement of the cluster mass function vs. redshift constrains cosmological parameters, including those associated with dark energy \citep{cosmo_DEprobe_wang,cosmo_DEprobe_haiman,RCSconstraints,Rozo:2007yt,mantz07}. This has motivated the design of new wide-area cluster surveys in the optical \citep{DES}, X-ray \citep{ConX}, and using the Sunyaev-Zel'dovich effect (SZE) \citep{SPT_1,SPT_2}. The utility of this probe hinges on limiting the uncertainty in the relation between cluster mass and whatever observable (e.g., optical richness, X-ray luminosity, or SZE flux decrement) is used as a proxy for it. Measurement of the two-point correlation function of clusters can help calibrate such mass-observable relations and thereby improve the resulting cosmological constraints \citep{maj, lima}.

Since clusters are the largest virialized mass concentrations in the Universe, measurement of their spatial clustering also provides insight into models of large-scale structure formation and tests theoretical frameworks, such as the Halo Model, that describe the relation between the galaxy, cluster, and dark matter distributions, i.e., the bias \citep{RDF,bahcall_sim,SSS2006,Schulz2006,Smith2007}.  

On very large scales, $r \sim 100\Mpc$, the two-point correlation function or power spectrum of clusters should show evidence of baryon acoustic oscillations (BAO) \citep{Angulo2005}. In concert with measurements of the cosmic microwave background anisotropy, the BAO scale provides an estimate of cosmic distance and thereby a geometric probe of dark energy \citep{Seo2003,hu03}. A possible detection of BAO in the power spectrum of clusters from the Abell/ACO catalog was reported in  \cite{mnb}, while earlier studies had claimed evidence for a feature in the cluster correlation function at $r \sim 125\Mpc$ \citep{kopylov,mo,einasto1,einasto2,einasto3}.

The BAO feature was detected at $3.4\sigma$ significance in the two-point correlation function of $\sim 47,000$ luminous red galaxies (LRGs) with spectroscopic redshifts in the range $0.16<z<0.47$ in the Sloan Digital Sky
Survey (SDSS) \citep{Eisenstein2005}. Using slightly larger samples, the BAO feature was also detected in the LRG power spectrum \citep{Huetsi2006,Tegmark2006,Percival2006}. 

The BAO feature has also been inferred from the large-scale clustering of $\sim 600,000$ LRGs identified in the deeper SDSS photometric survey \citep{Blake2007,Padmanabhan2007}. Although the photometric catalog covers a larger volume and contains many more galaxies than the spectroscopic sample,
the galaxy redshifts in the former must be estimated photometrically. With a photometric redshift uncertainty of $\sigma_z \sim 0.03$ for this sample, much
of the information in the radial modes of the power spectrum is lost, and the BAO feature was detected at less than $3\sigma$ significance. Based on theoretical considerations presented in \cite{Seo2003} and \cite{Blake2005}, \cite{Blake2007} presented a direct measurement of the 3D power spectrum, with a correction for the damping of power in the radial direction due to photometric redshift uncertainties. By contrast, \cite{Padmanabhan2007} measured the angular power spectrum in photometric redshift slices and used it to reconstruct the three-dimensional power spectrum. 

In this paper, we present measurements of the two-point correlation function for optically selected galaxy clusters in the Sloan Digital Sky Survey (SDSS) and study the possible detection of baryonic oscillations. The cluster samples are derived from the MaxBCG catalog \citep{maxBCGsample,Koester2007}, in which clusters are identified as concentrations of red-sequence galaxies; the colors of these early-type galaxies are used to estimate photometric redshifts for the clusters. The two point correlation function for
galaxy clusters in SDSS was studied in \cite{Basilakos} using an earlier cluster catalog.

We focus on measurement of the 3D cluster correlation function. In comparing the observations to models of large-scale structure, the model predictions are corrected for the effects of photometric redshift errors. We use two different methods for estimating this correction which are in good agreement for the photometric redshift error $\sigma_z \sim 0.01$ characteristic of the MaxBCG catalog. 

The cluster correlation function measurements provide weak evidence ($\sim 1.4-1.7\sigma$) for the presence of BAO in the cluster spatial distribution. An independent measurement of the power spectrum for the same cluster sample has been presented in \citep{Huetsi2007}, also indicating weak evidence ($\sim 2\sigma$) for acoustic features in the power spectrum.

The paper is organized as follows. In section \ref{sec:maxbcg_cf} we describe the MaxBCG cluster catalog, the samples we derive from it, and our measurements of the cluster correlation function. In section \ref{sec:model} we introduce our model for the cluster correlation function and discuss the corrections due to photometric redshift errors, presenting two different correction methods. In section \ref{sec:covariance} two different estimates of the correlation function covariance matrix are described and compared. In section \ref{sec:results} we compare the model to the data, extracting estimates of the cluster bias as a function of richness and mass, and comparing the goodness of fit for models with and without BAO. We present our conclusions in section \ref{sec:conclusions}.

\section{The MaxBCG cluster correlation functions}
\label{sec:maxbcg_cf}

\subsection{The MaxBCG cluster catalog}
\label{subsec:cat}

The cluster samples we analyze are derived from the SDSS MaxBCG catalog \citep{maxBCGsample} for SDSS DR5 \citep{DR5}. The MaxBCG method \citep{Koester2007} identifies clusters using two optical properties. First, the 
brightest cluster galaxy (BCG) typically lies near the center of the cluster 
galaxy light distribution. Second, the cores of rich clusters are dominated by 
red-sequence galaxies that occupy a narrow locus in color-magnitude space, the E/S0 ridge-line. MaxBCG uses a maximum-likelihood method to evaluate the probability that a given galaxy is a BCG near the center of a red-sequence galaxy density excess. Once a list of potential cluster centers is obtained, galaxies are grouped around those centers and the clusters are identified.

One measure of the richness of the clusters, denoted $\Ng$, is defined as the number of galaxies on the E/SO red sequence brighter than $0.4L^*$ that lie within a scaled radius $R_{200}$ of the BCG, where $R_{200}$ is the radius within which the density of galaxies with $-24<M_r<-16$ is 200 times the mean density of such galaxies \citep{hansen}. Dynamical \citep{Becker:2007iv} and statistical weak lensing measurements \citep{Johnston2007,maxBCG_weaklensing1}
indicate that $\Ng$ is strongly correlated with cluster virial mass. The public MaxBCG catalog contains 13,823 clusters with $\Ng \geq 10$; the catalog is approximately volume-limited over the redshift range $0.1-0.3$ and covers 7500 square degrees. Tests on mock catalogs indicate that the MaxBCG sample should be $\gtrsim 90$\% pure and complete for clusters with $\Ng \geq 10$ \citep{maxBCGsample,RozoEtal2007cosmo}.   

We subdivide the MaxBCG catalog into four samples for analysis, using the following thresholds in cluster richness: $\Ng\ge 10$, $11$, $13$, and $16$. The corresponding virial mass thresholds, based on statistical weak lensing measurements, are approximately $4.3$, $5.1$, $6.1$, and $7.9 \times 10^{13}~h^{-1} M_\odot$ \citep{Johnston2007}. The resulting numbers of clusters $N_c$ and the spatial number densities $n_c$ (assuming a survey volume of $0.5 h^{-3}$ Gpc$^3$) for each sample are given in Table~\ref{tab:samples}. Note that by {\it virial mass} we 
mean the mass denoted by $M_{vir}$ in \citep{Johnston2007}; it is defined in terms of the overdensity at collapse with respect to the background density by the redshift- and cosmology-dependent formula, $\Delta_{vir}=(18\pi^2+82 x-39x^2)/(1+x)$, with $x=\Omega_m(z)-1$, derived in \citep{BryanNorman1998} for $\Lambda$CDM models. This definition corresponds to the mass $M$ that appears in the halo bias formulas of Section~\ref{subsec:clusterbias}, where we drop the subscript for simplicity.

\clearpage
\begin{table}[b]
\caption{\label{tab:samples} Cluster samples used in this analysis: 
$N_c$ is the number of clusters in each sample, $n_c$ indicates the 
mean number density for each sample, assuming a volume of $0.5\cGpc$.}
\begin{tabular}{lcc}
    sample         & $N_c$   & $n_c~[\icMpc]$      \\ 
\hline
$\Ng \ge 10$       & 13823  & $2.8\times 10^{-5}$ \\
$\Ng \ge 11$       & 11265  & $2.3\times 10^{-5}$ \\
$\Ng \ge 13$       & 7796   & $1.6\times 10^{-5}$ \\
$\Ng \ge 16$       & 4853   & $1.0\times 10^{-5}$ \\
\end{tabular}
\end{table}
\clearpage

The redshifts for the MaxBCG clusters are estimated photometrically from the 
$g-r$ colors of the red-sequence galaxies. Since the color locus of these 
galaxies has finite width, the cluster photometric redshifts have a non-zero 
dispersion around the true values. Tests using a subsample of the BCGs with 
spectroscopic redshifts indicate that the dispersion in the cluster photometric 
redshift estimates is approximately $\sigma_z \equiv \langle (z_{\rm ph}-z_{\rm sp})^2 \rangle^{1/2}\simeq 0.01$, with a small dependence on richness \citep{maxBCGsample}. For the $\Ng \geq 10$ sample, the dispersion varies from $0.006$ at the lower end of the redshift range ($z=0.1$) to $0.011$ at the upper end ($z=0.3$). The error distribution is generally well described by a Gaussian \citep{maxBCGsample}. 

As we will show, for separations larger than $\sim 50\Mpc$, the translation of 
the cluster correlation function from real space to photometric-redshift 
space depends sensitively on the photometric redshift uncertainty, $\sigma_z$. In this work, 
for simplicity we assume a constant photo-z dispersion of $\sigma_z=0.01$, as suggested by 
\citep{maxBCGsample}, as our default. Throughout the paper, and particularly in 
\S \ref{subsubsec:compare}, \ref{subsec:estimate} and \ref{subsubsec:photozresults}, 
we discuss the systematic errors associated with the 
uncertainty in $\sigma_z$.

\subsection{Estimation of the two-point correlation function}
\label{subsec:est}

We measure the correlation function $\xi(r)$ for each cluster sample by means
of the Landy-Szalay estimator \citep{LS1993},
\bea
1+\xi_{LS}(r) & = & \frac{1}{RR(r)}\times \nonumber \\
& & \!\!\!\left[DD(r)\frac{n_R^2}{n_D^2}-2DR(r)\frac{n_R}{n_D}+RR(r)\right] 
\label{eq:LS}
\eea
where $DD(r)$ represents the number of cluster pairs with separation $r\pm 
\Delta r/2$ in the data, $RR(r)$ is the number of pairs in the same separation 
bin in a random catalog  uniformly distributed over the same survey volume, 
$DR(r)$ is the number of pairs with one member of the pair from the data 
sample and the other from the random catalog, $n_R$ is the number density of 
clusters in the random catalog, and $n_D$ is the number density of the data catalog. To reduce the effects of shot noise, the random catalog is five times denser than the data catalog. The random catalog is generated using the same angular mask used for the weak lensing analysis of the MaxBCG catalog 
\citep{maxBCG_weaklensing1} and with the redshift distribution measured in the 
data, see Fig.~4 in \cite{maxBCGsample}. 
 
Due to selection effects, the redshift distribution changes slightly as a 
function of the richness threshold, so a different random catalog is generated 
for each sample. The separation $r$ is given in comoving coordinates, and it is 
obtained assuming a flat $\Lambda$CDM cosmology with $\Omega_m = 0.27$. 
Where needed, we adopt the following other cosmological parameter values in 
this work: the Hubble parameter $h \equiv H_0/100 {\rm km/s/Mpc}=0.72$, the 
baryon density $\Omega_b=0.046$, the primordial perturbation spectral index 
$n_s=1$, and linear power spectrum amplitude $\sigma_8=0.9$.

\section{A model for the cluster correlation function}
\label{sec:model}

In this section, we develop a model for the cluster correlation function that
includes the effects of non-linear evolution and bias (\S \ref{subsec:nonlin}) 
and photometric redshift errors (\S \ref{subsec:photoz}) and that can be 
compared to the measurement of the correlation function on large scales, 
$r \sim 20 - 200 ~h^{-1}$ Mpc, with particular attention to the region where 
the BAO feature is expected.

\subsection{Non-linear evolution and bias}
\label{subsec:nonlin}

Although the rms density perturbation amplitude on large scales is much smaller than unity, one cannot rely on linear perturbation theory to precisely predict the large-scale correlation function, especially on the BAO scale of $\sim 100$ Mpc. The effects of non-linear evolution of perturbations on the acoustic features in the matter power spectrum and correlation function have been the subject of several studies in the recent literature, using analytic and semi-analytic techniques and N-body simulations \citep{Meiksin1999,Seo2003,White2005,ESSS2006,Jeong2006,Guzik2007,Huff2007,
Crocce2007,Matarrese2007,Smith2007,Angulo2007}. It has been recognized that the growth of structure induces a substantial damping of the acoustic peak in the correlation function with respect to linear theory; this damping must be taken into account when comparing theory with observations. 

In this work, we model this effect using Renormalized Perturbation Theory (RPT) \citep{Crocce2006,Crocce2007}. This prescription has the advantage of being based exclusively on first principles and, as shown in \citep{Crocce2007}, achieves remarkable agreement with results from N-body simulations. In general, the RPT non-linear matter power spectrum can be expressed as the sum of a term that accounts for the degradation of the initial power at a given wavenumber $k$ and a term arising from the non-linear gravitational coupling of modes of different wavenumbers,
\beq
P_{RPT}(k)=G^2(k;a) P_L(k)+P_{MC}(k;a)~.
\label{eq:RPT}
\eeq
Here $G(k;a)$ is the RPT propagator from the initial conditions, $a$ is the cosmic scale factor, $P_L$ is the power spectrum in linear perturbation theory for the $\Lambda$CDM model, and $P_{MC}$ is the contribution from mode coupling. We refer the reader to  \cite{Crocce2006} and \cite{Crocce2007} for detailed descriptions of the computation of the power spectrum in RPT. As shown in \citep{Crocce2007}, the first term in Eqn.~(\ref{eq:RPT}) is primarily responsible for the damping of the acoustic peak in the correlation function, while the second term gives a small correction of the order of a few percent at the relevant scales. We neglect the latter term in this analysis, since its contribution is subdominant compared to the observational errors in our measurements. 

While the RPT model has been validated against simulations around the BAO scale, 
its validity on the smallest scales we consider in our analysis, $r \sim 20 ~h^{-1}$ Mpc, has not yet been properly tested. We therefore limit its use to describing the damping of the baryonic peak at large scales. To model the non-linear evolution of the non-BAO part of the  power spectrum amplitude on small scales we use the \texttt{halofit} code \citep{Smith2003}. As a result, our adopted non-linear matter power spectrum model is given by 
\bea\label{eq:PNL}
P_{NL}(k) & = & [P_L(k)-P_{L,nw}(k)]G^2(k;a)\nonumber\\
& & +P_{{\rm HF},nw}(k;a)~,
\eea
where the ``no-wiggles'' $P_{L,nw}$ and $P_{HF,nw}$ are the linear power spectrum and non-linear \texttt{halofit} power spectrum with the acoustic oscillations edited out using the featureless transfer function derived in \citep{EisensteinHu1998}. Given the measurement errors for the current cluster sample and the uncertainties in the photo-z errors (\S \ref{subsec:photoz}), we find that the theoretical uncertainties in modeling the non-linear evolution 
of the baryonic peak---as reflected in the different approaches in the 
recent literature---are negligible by comparison.  For example, we find that using the alternative prescription for the non-linear matter power spectrum 
$P_{NL}(k)$ followed by \citep{Eisenstein2005} does not change the significance of the BAO feature in this work. 

The dark matter correlation function is obtained in the usual way via the Fourier transform of the non-linear matter power spectrum, 
\beq\label{eq:xil}
\xi_{mm}(r)=\frac{1}{2\pi^2}\int P_{NL}(k) \frac{\sin{kr}}{kr}k^2 {\rm d}r ~.
\eeq
For the $\Lambda$CDM model parameters given in \S \ref{subsec:est}, the predicted non-linear correlation function is shown as the black dashed curve in 
Fig.~\ref{fig_cf_photoz} below. To connect this to observations, we will assume, for simplicity, that the correlation function for a given cluster sample is related to the matter correlation function by a constant bias factor, 
\beq
\xi_{cc}(r)=b^2~\xi_{mm}(r)~.
\eeq
A more accurate description of cluster bias is possible, but again the 
uncertainties in the bias prescription are expected to be small compared to the 
current measurement errors. We treat the bias factor $b$ as a fit parameter in 
comparing the model to observations; in \S \ref{subsec:clusterbias} we compare 
the fit values of the bias with predictions from the Halo Model.

\subsection{Effects of photometric redshift errors}
\label{subsec:photoz}

As noted in \S \ref{subsec:cat}, the estimated photometric redshifts 
(photo-$z$'s) for the MaxBCG clusters have a non-negligible uncertainty, 
$\sigma_z \simeq 0.01$. This translates into a positional uncertainty along the 
line-of-sight of approximately $30\Mpc$, depending on cosmological parameters. 
The effect on the three-dimensional correlation function is a smearing of 
the acoustic peak and a relative damping of power on small scales.
We consider and compare two  prescriptions to model this effect, one analytic, the other based on a direct geometric approach. For simplicity, we ignore the 
redshift dependence of $\sigma_z$ in this analysis. We note that the effect 
of photo-z errors is analogous to but simpler to model than redshift-space distortions in spectroscopic surveys.

\subsubsection{Analytic power spectrum smearing}

In the first approach, we adopt the simple analytic prescription for power 
spectrum smearing introduced in \cite{Blake2005}. Assuming a Gaussian 
smearing along the line of sight due to photo-$z$ errors, the damping of the 
power spectrum in the plane-parallel approximation is given by  
\beq\label{ps_aniso}
P_{c, \sigma}(k_\perp,k_z)=P_{c}(k)e^{-k_z^2\sigma^2}~,
\eeq
where $k_\perp$ and $k_z$ are the components of the wavevector $\kv$ 
perpendicular and parallel to the line of sight, $P_{c}(k)=b^2P_{NL}(k)$ is the 
true (real-space) non-linear cluster power spectrum, and $\sigma$ is the 
dispersion in comoving distance along the line of sight, which is related to the 
photometric redshift error $\sigma_z$ by 
\beq
\sigma = \frac{c}{H(z)}\sigma_z~.
\eeq
In this expression, we evaluate the Hubble parameter at the effective median 
sample redshift $z=0.22$. Although the photo-$z$ correction to the power spectrum is clearly anisotropic in Fourier space, one can retrieve the monopole part of the observed (photo-$z$ space) power spectrum as 
\bea
\label{eq:ps_iso}
\widetilde{P}_{c}(k) & \equiv &
\frac{1}{2}\int d\cos\theta~ P_{c, \sigma}(k\sin\theta,k\cos\theta) \nonumber \\
& = & \frac{\sqrt{\pi}}{2\sigma k} {\rm erf}(\sigma k) 
P_{c}(k) ~,
\eea
where $\theta$ is the angle between $\kv$ and the line of sight (see also 
\citep{Huetsi2007}). The monopole of the measured (photo-$z$-space) correlation 
function can then be computed as the Fourier transform of 
$\widetilde{P}_{c}(k)$.

\subsubsection{Geometric smearing}

\begin{figure}[t]
\begin{center}
\includegraphics[width=0.98\columnwidth]{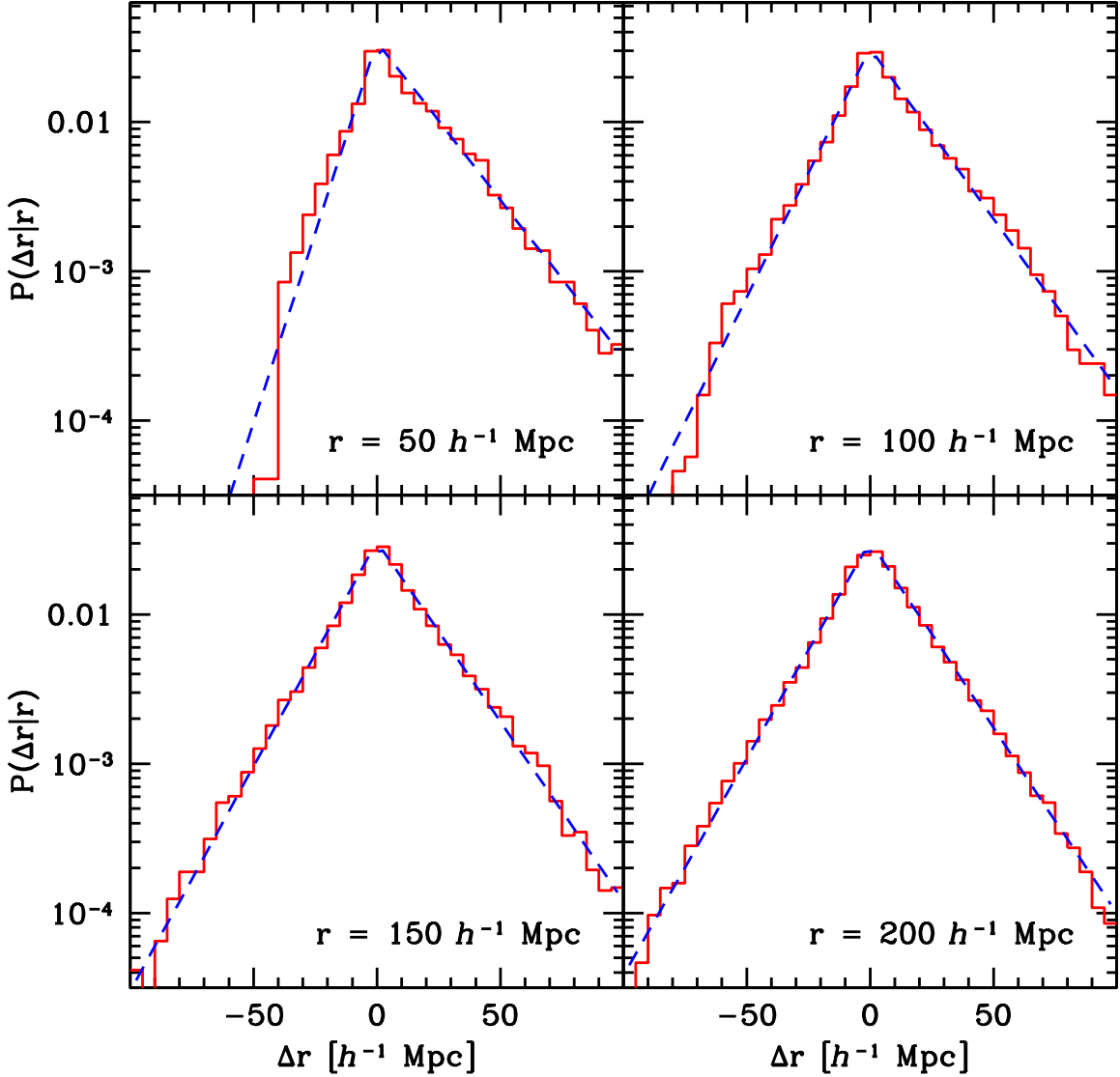}
\caption{\label{fig:histograms}
Normalized, conditional probability distributions $P(\Delta r\,|\,r)$ of the  
difference $\Delta r = \tr-r$, given $r = 50$, $100$, $150$, and $200\Mpc$, as 
measured from the random catalogs, assuming Gaussian photo-$z$ errors with 
$\sigma_z=0.01$. The dashed blue lines correspond to the exponential fit 
of eq.~(\ref{eq:laplace}).}
\end{center}
\end{figure}

While convenient, the analytic approach above does not take into account light-cone effects or the effects of the survey geometry. Here we directly model the geometric effects of photo-$z$ errors on the measurement of the correlation function. For simplicity, we consider a random Poisson distribution of points covering the survey volume, using the same random catalog that is used in the two-point function estimator (\S \ref{subsec:est}). We displace each point along the line of sight by a random distance drawn from a Gaussian probability distribution, 
\beq
P(z_{\rm ph})\sim \exp\left[-\frac{(z_{\rm ph}-z)^2}{2\sigma^2_z}\right] ~,
\eeq
where $z_{\rm ph}$ is the simulated photometric redshift estimate for a 
point at true redshift $z$, and $\sigma_z$ is the standard deviation 
of the photo-$z$ estimate. For each pair with true separation $r$ in the random catalog, we obtain a ``measured'' separation $\tr$ after the displacements due to photo-$z$ errors. We then calculate the conditional probability $P(\tr|r)$ for a pair to have a measured separation $\tr$ after displacement, given the true separation $r$, or, equivalently the probability $P(\Delta r|r)$ for the difference $\Delta r \equiv \tr - r$, given the value of $r$. In Fig.~\ref{fig:histograms} we plot the normalized probability distributions for $\Delta r$ corresponding to four representative values of $r$.

\clearpage
\clearpage

\begin{figure}[t]
\begin{center}
\includegraphics[width=0.98\columnwidth]{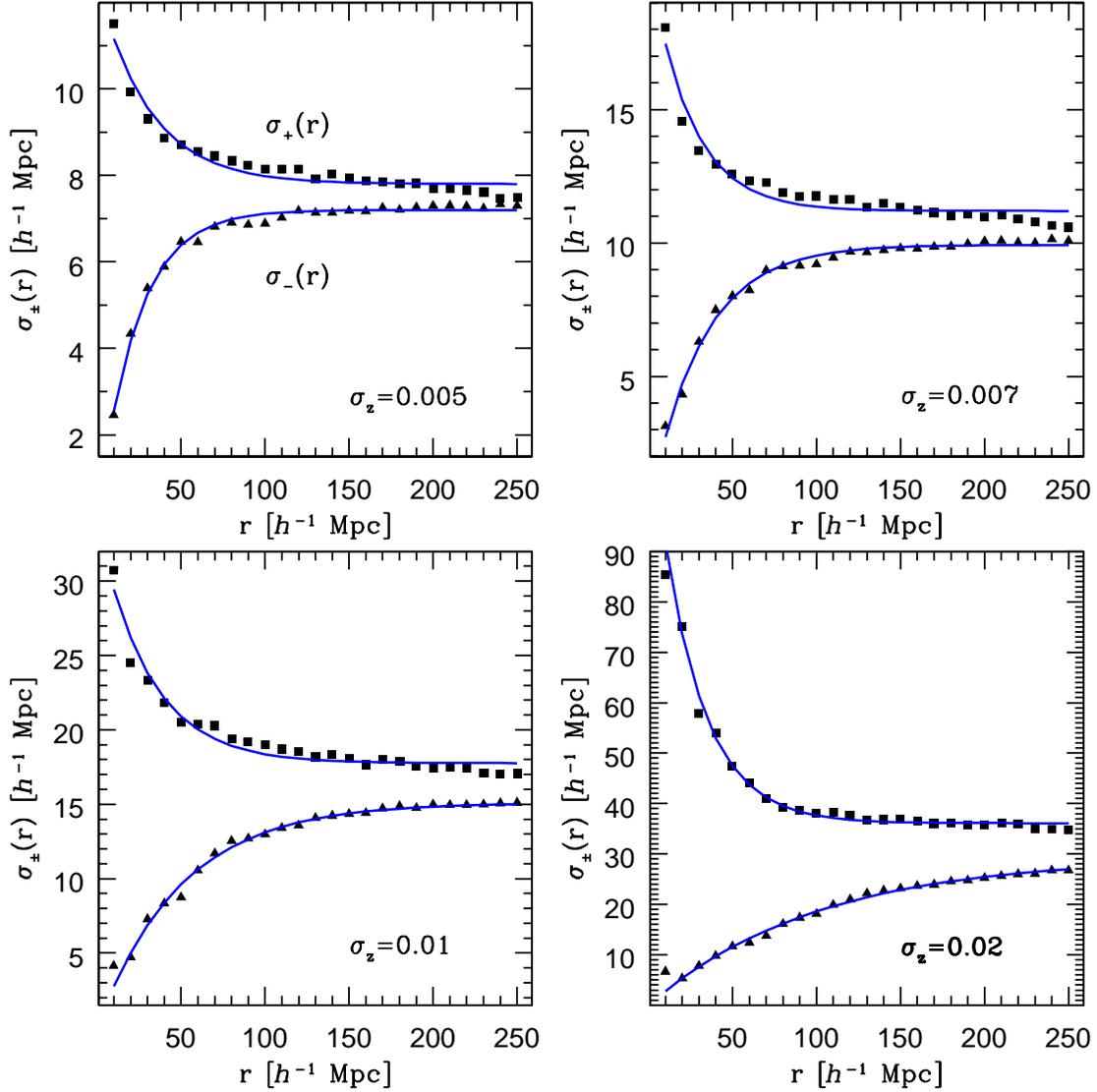}
\caption{\label{fig:sigmaR} Values of $\sigma_+(r)$ and $\sigma_-(r)$ measured from the Poisson realizations and corresponding fitting functions from Eqns. (\ref{eqn:sigp}) and (\ref{eqn:sigm}) as determined from the measured probability distributions $P(\Delta r|r)$ for a Gaussian photometric-redshift error of $\sigma_z=0.005$, $0.007$, $0.01$ and $0.02$.} 
\end{center}
\end{figure}

By definition, $\Delta r$ is limited by $\Delta r \ge -r$, and we therefore 
expect $P(\Delta r|r)$ to be asymmetric in $\Delta r$. As 
Fig.~\ref{fig:histograms} suggests, $P(\Delta r|r)$ is well-described by an 
asymmetric Laplace (or exponential) distribution,
\beq\label{eq:laplace}
P(\Delta r|r)=C(r)\times\left\{
\begin{array}{cc}
\exp{\left[-\frac{\Delta r}{\sigma_{+}(r)}\right]} 
&  \quad {\rm for}~ \Delta r \ge 0 \\
\exp{\left[\frac{\Delta r}{\sigma_{-}(r)}\right]} 
& \quad {\rm for}~ \Delta r < 0 
\end{array}\right.
\eeq
which is continuous at $\Delta r=0$, and where the normalization factor is 
\beq
C(r)=\frac{1}{\sigma_+(r)+\sigma_-(r)[1-\exp(-r/\sigma_-(r))]}.
\eeq
We fit the functional form in Eqn.~(\ref{eq:laplace}) to the probability distributions measured for values of $r$ ranging from $10$ to $250\Mpc$ in steps of $10\Mpc$. We then fit the function $\sigma_{+}(r)$ to an exponential functional form with three parameters, 
\beq
\sigma_+(r)=\sigma_{+,0}\left(1+c_{+} e^{-r/r_{+}}\right),
\label{eqn:sigp}
\eeq
while for $\sigma_-(r)$, imposing the condition $\sigma_-(0)=0$, we consider the two-parameter form
\beq
\sigma_-(r)=\sigma_{-,0}\left(1- e^{-r/r_-}\right) ~.
\label{eqn:sigm}
\eeq
The resulting parameter values, for four different values of the photo-z dispersion, $\sigma_z=0.005$, $0.007$, $0.01$ and $0.02$, are given in Table~\ref{tab:sigma_fits}. These functions, together with the values measured from the realizations, are shown in Fig.~\ref{fig:sigmaR}. Note that the probability distribution is never symmetric, as the two functions $\sigma_{+}(r)$ and $\sigma_{-}(r)$ differ significantly at all relevant scales. 

\clearpage
\begin{table}[b]
\caption{\label{tab:sigma_fits} 
Values of the parameters determined by fitting the exponential expressions for 
$\sigma_{\pm}(r)$ to the dispersions measured from the Poisson realizations
corresponding to $\sigma_z=0.005$, $0.007$, $0.01$ and $0.02$.}
\begin{tabular}{l|cc|ccc}
\hline
$\sigma_z$     
        &\multicolumn{2}{c}{$\sigma_-(r)$}
                               &\multicolumn{3}{c}{$\sigma_+(r)$}\\
        &$\sigma_{-,0}$&$r_-$  &$\sigma_{+,0}$&$r_+$ &$c_+$      \\
\hline
$0.005$ & $7.20$       &$22.9$ & $7.80$       &$30.9$&$0.596$    \\ 
$0.007$ & $9.93$       &$31.1$ & $11.2$       &$24.6$&$0.840$    \\ 
$0.010$ & $15.1$       &$49.4$ & $17.8$       &$30.5$&$0.913$    \\ 
$0.020$ & $29.4$       &$100.0$& $36.1$       &$25.0$&$2.318$    \\ 
\end{tabular}
\end{table}
\clearpage

\clearpage

Using this procedure, we obtain an analytic expression for the probability function $P(\Delta r|r)$. The measured (photo-$z$-space) cluster correlation function, $\tilde{\xi}_{cc}(\tr)$, can then be approximately derived from the real-space correlation function $\xi(r)$ by a normalized convolution, 
\beq\label{eq:xi_ph}
\tilde{\xi}_{cc}(\tr) = 
\frac{\int_0^\infty \xi_{cc}(r) P(\tilde{r}|r)RR(r) {\rm d}r}
{\int_0^\infty P(\tilde{r}|r) RR(r) {\rm d}r} ~,
\eeq
where $P(\tilde{r}|r)\equiv P(\Delta r|r)$, and $RR(r)$ is the total number of pairs in the random catalog at the actual separation $r$. We find that the separation dependence of this quantity in the random catalog can be fit over the relevant range of scales by 
\beq
RR(r) = \left(\frac{r}{1.35}\right)^2-\left(\frac{r}{9.11}\right)^3+
\left(\frac{r}{32.0}\right)^4 ~,
\eeq
where the separation $r$ is in units of $h^{-1}$ Mpc.

\clearpage
\begin{figure*}[p]
\begin{center}
\includegraphics[width=0.8\textwidth]{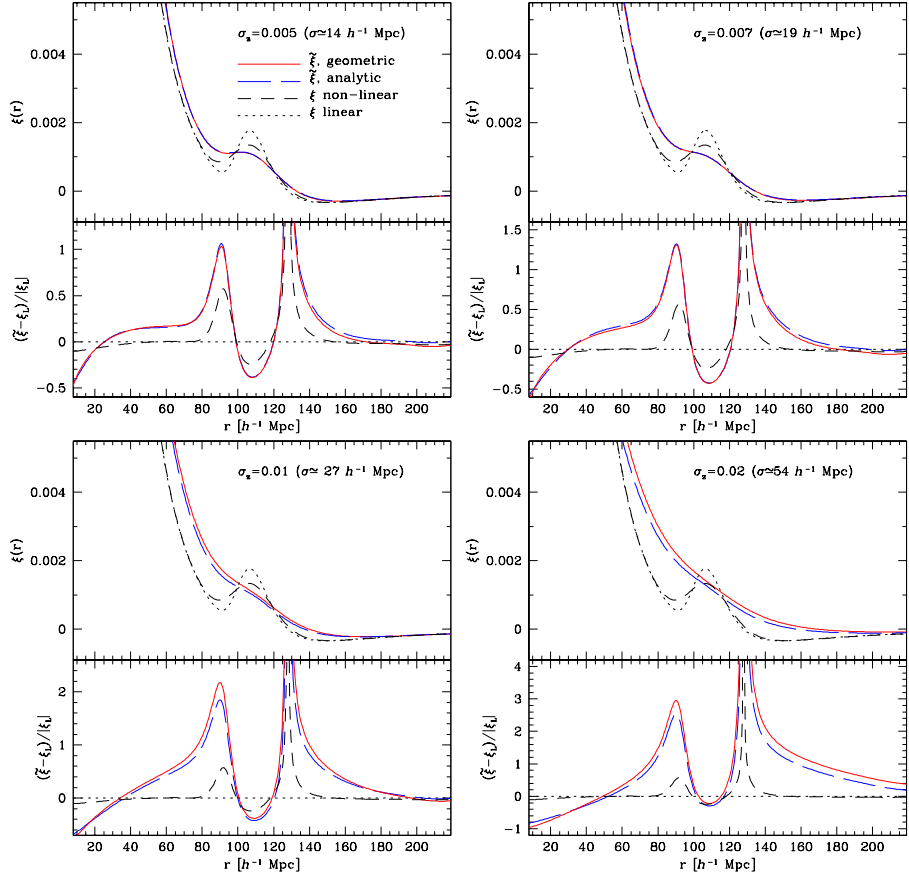}
\caption{\label{fig_cf_photoz} 
Predicted matter two-point correlation function, including correction for photo-$z$ errors. Dashed black curve shows the non-linear matter correlation function (Eqn.~\ref{eq:xil}) for the $\Lambda$CDM model, while the dotted black curve shows the linear theory correlation function. Long-dashed blue and solid red curves show the non-linear correlation function in photo-$z$ space for Gaussian 
photo-$z$ errors with $\sigma_z=0.005$, $0.007$, $0.01$, and $0.02$. Long-dashed blue curves use the analytic plane-parallel approximation of Eqn.~(\ref{eq:ps_iso}), solid red curves use the geometric convolution of Eqn.~(\ref{eq:xi_ph}). In all cases, the BAO feature is severely smoothed by the photo-$z$ errors. Corresponding curves in the lower parts of each panel show the relative differences with respect to the linear theory prediction; the spike around $r\simeq 130 \Mpc$ corresponds to the scale at which the linear correlation function vanishes.}
\end{center}
\end{figure*}
\clearpage

\subsubsection{Comparison}
\label{subsubsec:compare}

In Fig.~\ref{fig_cf_photoz} we compare these two methods of estimating the impact of photo-$z$ errors on the two-point correlation function, for Gaussian photo-$z$ errors with dispersion $\sigma_z=0.005$, $0.007$, $0.01$, and $0.02$. The continuous red curves show the photo-$z$-space correlation functions calculated using the geometric convolution in Eqn.~(\ref{eq:xi_ph}); the long-dashed, blue curves show the corresponding results using the Fourier transform of the analytic approximation of Eqn.~(\ref{eq:ps_iso}). On small scales, $r \lesssim 30~
h^{-1}$ Mpc, photo-$z$ errors strongly suppress the correlation function amplitude and flatten its 
slope. On larger scales, as the lower left panel of Fig.~\ref{fig_cf_photoz} shows, the combination of non-linearity and photo-$z$ error transforms the expected BAO bump into a more subtle inflection feature in the two-point correlation function. As a result, when statistical errors are included, we do not expect to be able to detect the BAO signature with high significance in this data set. 

The geometric and analytic results are in excellent agreement with each other for 
small values of the photo-$z$ dispersion, but a difference is observed for $\sigma_z\gtrsim 0.01$. 
In the data analysis and in the rest of the paper we adopt the geometric model and assume a 
constant error $\sigma_z=0.01$ for our fiducial results, 
except where otherwise noted. In this case, the photo-$z$ error changes 
the amplitude of the {\it non}-linear correlation function---in going from real to 
photo-$z$ space---by as much as $170\%$ on scales $r \sim 
90 ~h^{-1}$ Mpc. However, we 
note that even a small uncertainty in the photo-$z$ dispersion $\sigma_z$ introduces a significant 
systematic uncertainty in the mapping of the correlation function amplitude from real to 
photo-$z$ space, 
due to the difference in the smoothing of the acoustic features. 
For some of our results below, we therefore estimate a systematic error in the inferred 
real-space correlation function by 
comparing results from the geometric model for $\sigma_z=0.01$ with those for $\sigma_z=0.007$. 
At the same time, we note that a sufficiently large spectroscopic sample can in principle reduce 
the uncertainty in $\sigma_z$ to a small level. We plan to carry out a more detailed study of 
this and other issues in the modeling of the correlation function in photo-$z$ space 
elsewhere \citep{Estradaetal}.

\section{The covariance matrix}
\label{sec:covariance}

In order to compare the model predictions to the data and extract parameter measurements, we must have an estimate of the error covariance for the correlation function. There are two common procedures for estimating the errors. The first uses a jackknife estimator by creating subsamples from the data; this has the advantage of being independent of model assumptions, but it may not properly account for the variance due to modes on scales larger than those spanned by the survey. The second estimates the errors from a model, either using the variance among a large number of survey mock catalogs or, on large scales, using an analytic estimate assuming Gaussian perturbations. The model errors in principle account for cosmic variance due to modes on arbitrarily large scales, but they assume the model provides an accurate representation of the data. In the data-model comparison of \S \ref{sec:results}, we present fit results based on both the jackknife and an analytic prediction; in this section, we compute these error estimates and compare them.

\begin{figure}[t]
\begin{center}
\includegraphics[width=.98\columnwidth]{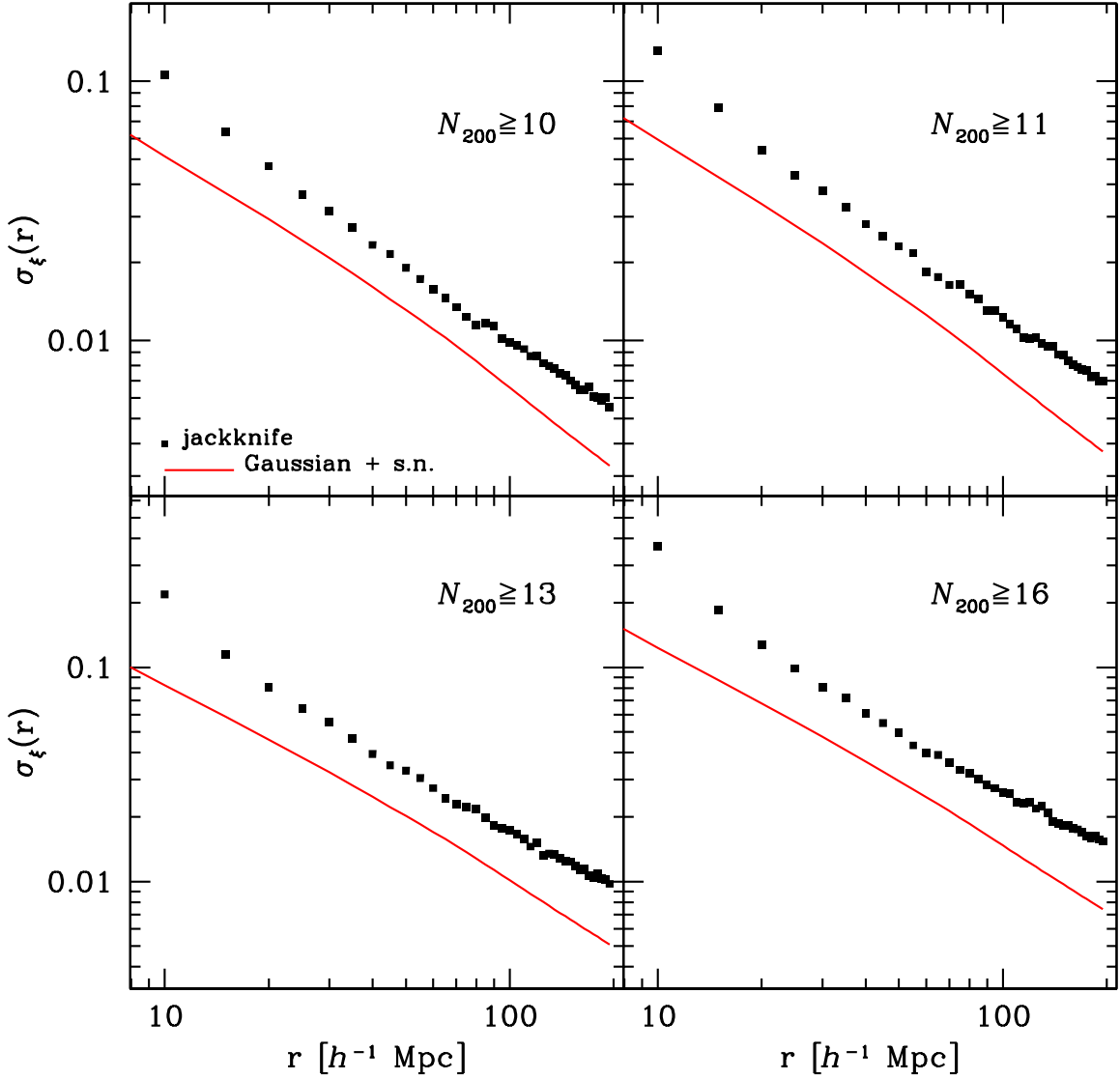}
\caption{\label{fig:variance} Standard deviations $\sigma_\xi(r_i) \equiv \sqrt{{\rm Cov}(\xi_i,\xi_ i)}$ of the correlation functions $\xi(r)$ for the four MaxBCG samples, estimated using the jackknife technique (black points) and as predicted by linear perturbation theory including the shot-noise contribution in Eqn.~(\ref{eq:covar_linear}) (solid red). The linear theory prediction assumes the $\Lambda$CDM model parameters given in \S \ref{subsec:est} and values of the linear bias parameter derived in 
\S \ref{sec:results}.}
\end{center}
\end{figure}

For the jackknife error estimate, we produced $1000$ subsamples of each of the 
four MaxBCG samples, in each case with $1/1000$th of the clusters removed at 
random. This procedure corresponds to the ``traditional'' jackknife approach in 
statistics, but it differs from the standard jackknife practice in large-scale structure studies, in which entire subvolumes are removed at random. The correlation function is measured for each subsample, and the covariance matrix is estimated from 
\beq\label{eq:covar}
{\rm Cov}(\xi_i,\xi_ j)=\frac{N-1}{N}
\sum_{l=1}^{N}(\xi_i^l-\bar{\xi_i})(\xi_j^l-\bar{\xi_j}) ~,
\eeq
where $\xi_i^l=\xi^l(r_i)$ is the correlation amplitude in the $i$th separation 
bin, $r_i$, for subsample $l$, and $N=1000$ is the number of subsamples. 
In Fig.~\ref{fig:variance} we show the diagonal jackknife standard deviation, 
$\sigma_\xi(r_i) \equiv \sqrt{{\rm Cov}(\xi_i,\xi_ i)}$, for the four samples 
(black points). As a test we have also performed jackknife measurements by removing 
subvolumes. When the number of removed subvolumes is small, the jackknife measurements 
are noisy; when it is large, the results appear to converge to those from 
the ``removal of clusters'' procedure above.

In the alternative approach to error estimation, 
the predicted error covariance assuming Gaussian perturbations is given by
\bea\label{eq:covar_linear}
{\rm Cov}(\xi_i,\xi_ j)& \simeq &
\frac{64\pi^4}{V}\int_0^\infty{\rm d} k k^2 P_{tot}^2(k)\nonumber\\
& & \times\frac{\sin(k r_i)}{k r_i}\frac{\sin(k r_j)}{k r_j} ~.
\eea
Here $V$ is the survey volume and 
\beq
\label{eq:Ptot}
P_{tot}(k)=b^2P_{NL}(k)+\frac{1}{n_{c}}\,,
\eeq
where $b$ is the linear bias parameter, $P_{NL}(k)$ is the non-linear matter 
power spectrum computed as described in the previous section,
and the second term in Eqn.~(\ref{eq:Ptot}) accounts for the shot-noise 
correction. For the term involving  
$1/n_{c}^2$ in Eqn.~(\ref{eq:covar_linear}) we make use of the integral
\bea
\int {\rm d} k \sin(k r_i) \sin(k r_j) & = & \frac{\pi}{2}\d_D(r_i-r_j)
\nonumber\\
 & \simeq &  \frac{\pi}{2~\Delta r}\d_{ij}\,,
\eea
where $\Delta r=5\Mpc$ is the bin-size for the correlation function measurement.
For a detailed discussion of these shot-noise effects, see 
\cite{Cohn2006}. 

Fig.~\ref{fig:variance} shows the analytic prediction for correlation function 
standard deviations including the shot-noise correction (red, continuous curves), 
using the $\Lambda$CDM power spectrum and the best-fit values for the bias 
parameters from the analysis in \S \ref{sec:results} (note that those bias 
parameter estimates are derived using the jackknife covariance matrix). The 
predictions are consistent with the jackknife estimates at the few to 30\% 
level for all samples.

\clearpage
\clearpage

The expression in Eqn.~(\ref{eq:covar_linear}) is the analytic estimate for the covariance of the correlation function in real space. To estimate the covariance for $\xi$ in photo-$z$ space, a na\"ive approach would be to replace the expression for $P_{tot}(k)$ in Eqn.~(\ref{eq:Ptot}) with a photo-$z$-corrected expression analogous to that in Eqn.~(\ref{eq:ps_iso}), i.e., 
\beq
\widetilde{P}_{tot}(k)=b^2P_{NL}(k)
\frac{\sqrt{\pi}}{2\sigma k}{\rm erf}(\sigma k) +\frac{1}{n_{c}}\,.
\label{eq:Ptotcorr}
\eeq
This would result in a predicted covariance generally lower than that in real space and more discrepant with the jackknife estimates, about $50\%$ lower than the latter for all samples. Such an expression would account for the smoothing induced by photometric errors along the line of sight, but it does not include the extra component due to the intrinsic randomness of photo-$z$ displacements. We postpone a more detailed discussion of the covariance of the correlation function in photo-$z$ space to future work \citep{Estradaetal}; here, we limit ourselves to the analytic estimate of Eqn.~(\ref{eq:covar_linear}). This expression does not take into account contributions due to the non-Gaussianity generated by gravitational instability and due to the anisotropic geometry of the survey. In these respects, it should provide a lower bound for the actual correlation function covariance. 

\clearpage
\begin{figure}[t]
\begin{center}
\includegraphics[width=.7\columnwidth]{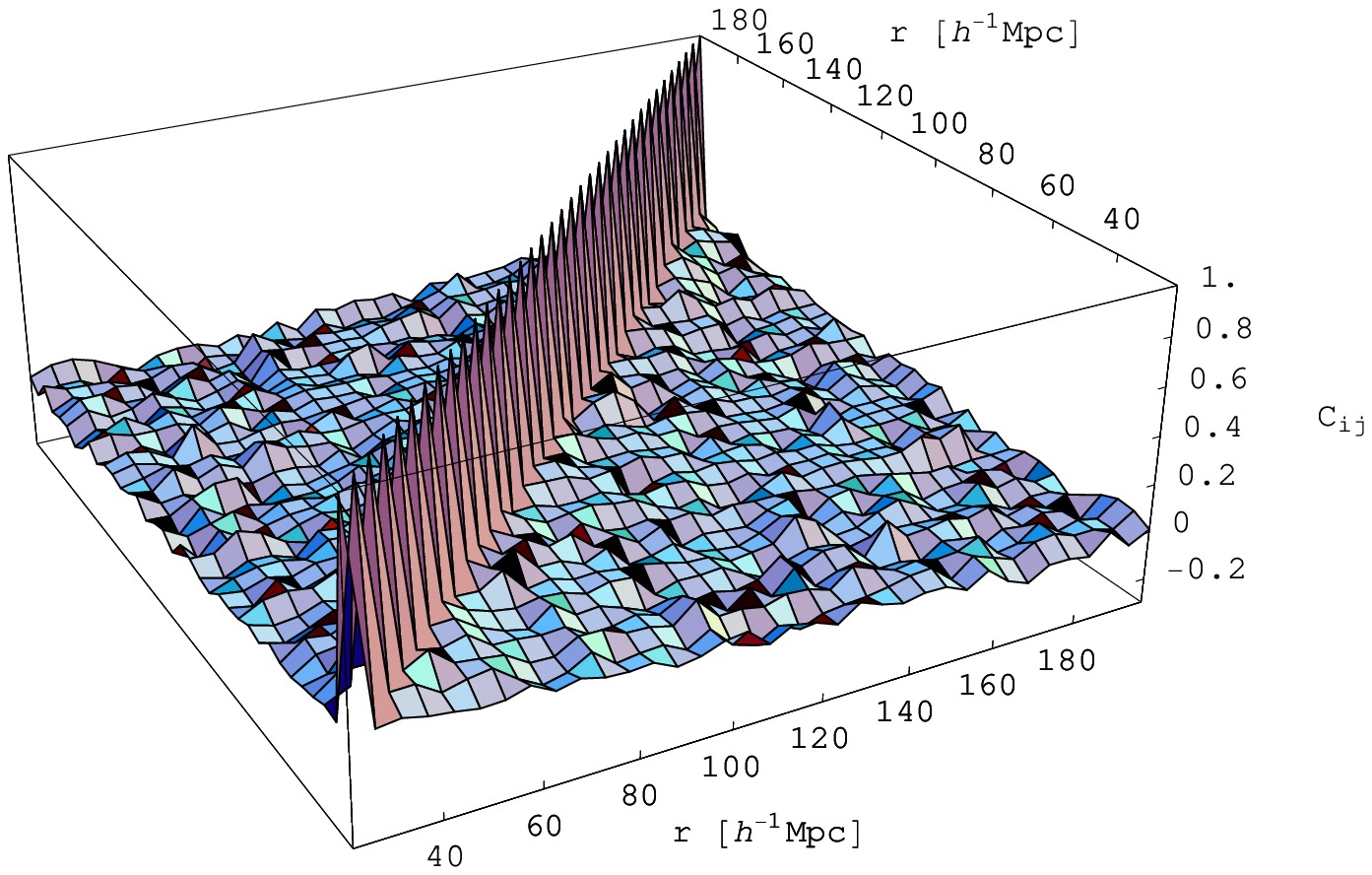}
\includegraphics[width=.7\columnwidth]{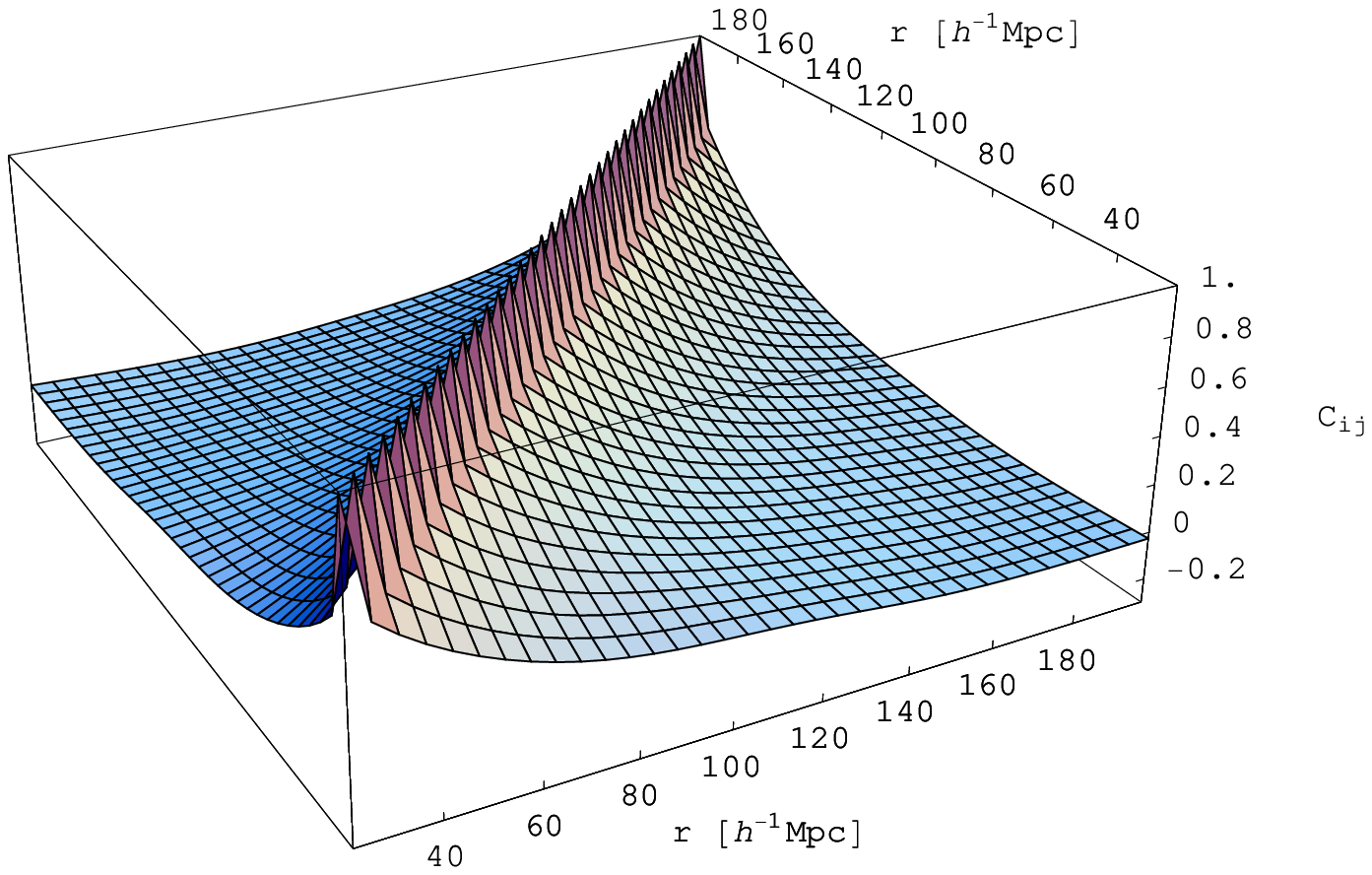}
\caption{\label{fig:covmatrix_N10} Cross-correlation coefficients for the binned correlation function for the $\Ng \ge 10$ sample determined from the jackknife technique (top) and from the Gaussian analytic prediction in real-space (bottom) including the shot-noise contribution. The horizontal axes show the linear bin separation label $i$, where $i=1$ corresponds to $r_i=5 \Mpc$ and $i=39$ to $r_i=195 \Mpc$.}
\end{center}
\end{figure}
\clearpage

In Fig.~\ref{fig:covmatrix_N10} we show the cross-correlation coefficients, defined by 
\beq\label{eq:redcovar}
C_{ij} \equiv \frac {{\rm Cov}(\xi_i,\xi_j)} 
{\sqrt{{\rm Cov}(\xi_i,\xi_i) {\rm Cov}(\xi_j,\xi_j)}}\,, 
\eeq
for the $\Ng \ge 10$ sample. We present results for both the jackknife estimator (top panel) and the Gaussian analytic prediction in real space (bottom panel) including the shot-noise contribution. The off-diagonal elements differ significantly between these two approaches, and it appears that the jackknife may underestimate them. Underestimating the off-diagonal elements could lead one to assign greater significance to features in the correlation function such as the acoustic peak, since it corresponds to underestimating the covariance between different separation bins. In other words, when comparing two models to the data, say, one with the BAO feature and the other without, if the jackknife and linear theory diagonal covariance elements are similar (as in, e.g., the $\Ng \ge 10$ case in Fig.~\ref{fig:variance}), then the jackknife error estimate will yield a larger value for the $\chi^2$ difference between the two models compared to the linear theory error estimate.

\section{Results}
\label{sec:results}

In this section we present the correlation function measured for the four cluster samples introduced in \S~\ref{sec:maxbcg_cf}, using the estimator of Eqn.(\ref{eq:LS}). We first present the measurement on scales $r \lesssim 60~h^{-1}$ Mpc and present fits for the correlation length and power-law slope of the two-point function, corrected for photo-$z$ errors. We then consider the correlation function on larger scales and compare to the model presented in 
\S~\ref{sec:model}, determining the best-fit values for the cluster linear bias parameter and the evidence for baryonic features.

\subsection{Estimate of the correlation length}
\label{subsec:estimate}

\clearpage
\begin{figure}[t]
\begin{center}
\includegraphics[width=.98\columnwidth]{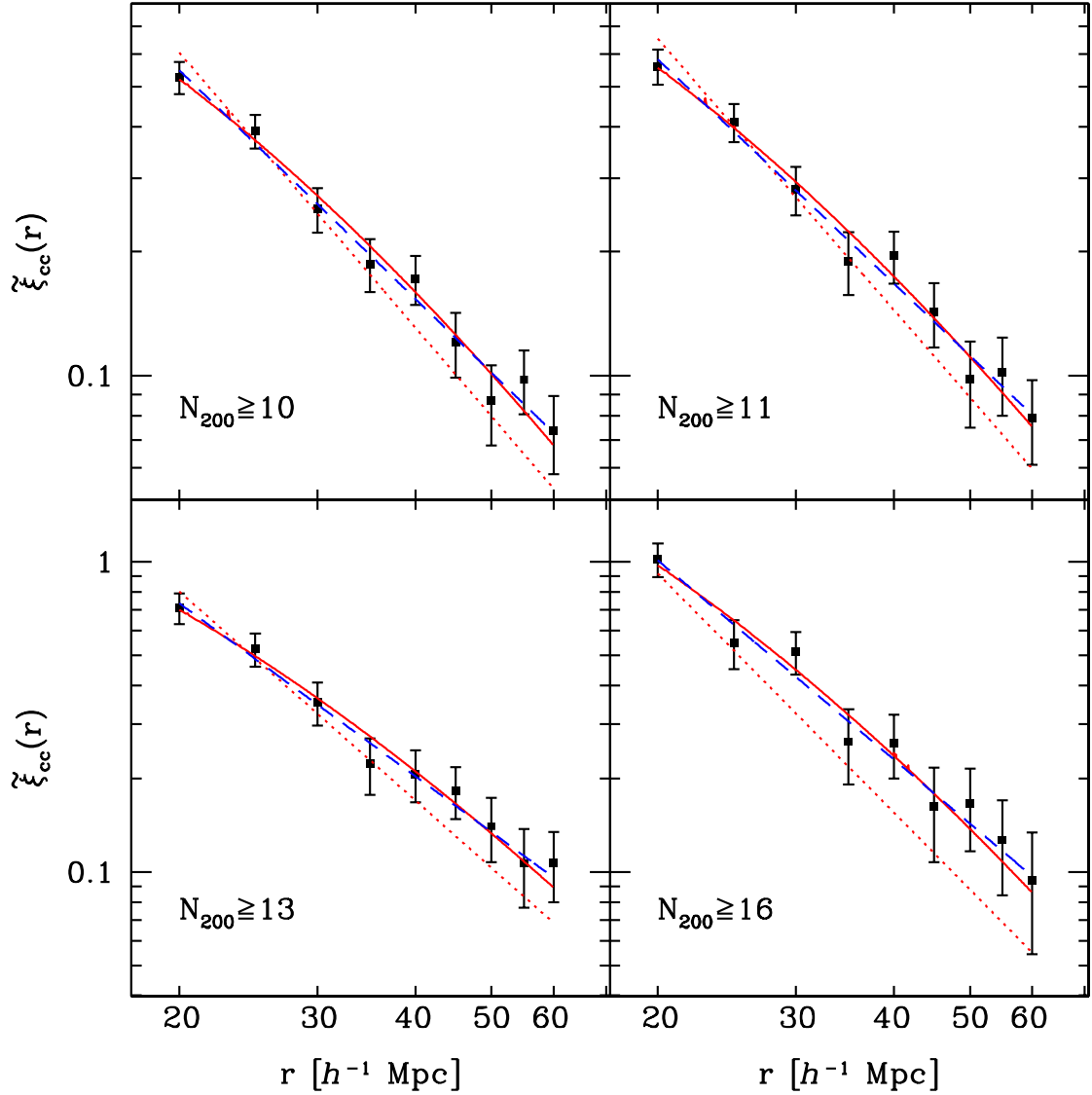}
\caption{\label{fig:powerlawfit} 
Data points show the cluster correlation function in photo-$z$ space measured for the four richness 
samples in bins of width $\Delta r = 5\Mpc$, with jackknife errors. Dotted 
red curves show inferred power-law $\xi_{cc}(r)$ in real space, assuming 
$\sigma_z=0.01$. Solid red curves show best-fit power-law models convolved 
with the photo-$z$ error distribution, ${\tilde \xi}_{cc}({\tilde r})$, which 
should match the data. Dashed blue curves show power-law fits to the data 
assuming no photo-$z$ error correction, $\sigma_z=0$.}  
\end{center}
\end{figure}
\clearpage

Historically, measurements of the cluster correlation function found results 
consistent with a power law over scales $r \lesssim 60~h^{-1}$ Mpc or so \citep{cc-meas_1,cc-meas_2,cc-meas_3,cc-meas_4,cc-meas_5},  
\beq\label{eq:powerlaw}
\xi_{cc}(r)= \left( \frac{r}{R_0} \right)^{-\gamma}\,,
\eeq
where the correlation length $R_0$ depends on cluster richness, and the slope is
$\gamma\sim 1.8$. A compilation of these results, along with measurements from 
an earlier, much smaller version of the MaxBCG catalog (from the SDSS Early 
Data Release, EDR), can be found in \cite{bahcall_data}. 

In Fig. \ref{fig:powerlawfit}, we show the estimated correlation functions for the four 
maxBCG samples over the range $20-60~h^{-1}$ Mpc. To fit these results to the 
power-law form of Eqn.~(\ref{eq:powerlaw}), we must include the effects of 
photometric redshift errors. We do 
this by inserting Eqn.~(\ref{eq:powerlaw}) into Eqn.~(\ref{eq:xi_ph}) 
and comparing with the data, assuming a photo-$z$ scatter of $\sigma_z=0.01$. 
The resulting best-fit correlation function in observable space, ${\tilde \xi}_{cc}({\tilde 
r})$, and the inferred power-law correlation function in real space, $\xi_{cc}(r)$, 
are shown as the solid and dotted red curves in Fig.~\ref{fig:powerlawfit}. 
In Table~\ref{tab:fitpowerlaw}, we present the inferred correlation length and 
slope in real space for the four MaxBCG samples. The statistical 
errors on $R_0$ and $\gamma$ come from the 
jackknife covariance. However, since there is some uncertainty 
in the photo-$z$ error variance, in the Table 
we show results for both $\sigma_z = 0.01$ and $0.007$; the difference between them 
provides an estimate for the systematic errors in the inferred parameters.  
We also show the inferred parameter values in the case that photo-$z$ errors are 
completely ignored, $\sigma_z=0$; these correspond to the power-law fit parameters 
in photo-$z$ space. 

\clearpage
\begin{table}[h]
\caption{\label{tab:fitpowerlaw} 
Power-law fits to the cluster correlation function on scales $r=20-60~h^{-1}$ Mpc, for three values of the photo-$z$ error variance, $\sigma_z=0.01$, 0.007, and 0. The inferred real-space correlation length $R_0$ is given in $\Mpc$, and $\gamma$ is the inferred slope of the real-space correlation function. The fourth  column gives the $\chi^2$ per degree of freedom for the best fit, while the fifth column shows the mean cluster separation $d$, in $\Mpc$, assuming a sample volume of $0.5\cGpc$.}
\begin{tabular}{l|ccc|c}
    sample    &$R_0$      & $\gamma$ & $\chi^2$/d.o.f. & $d$\\
\hline
\hline
\multicolumn{5}{l}{$\sigma_z=0.01$}\\
\hline
$\Ng \ge 10$  &$15.93\pm0.33$&$2.21\pm0.11$&$0.45$& $33.1$\\
$\Ng \ge 11$  &$16.45\pm0.38$&$2.18\pm0.12$&$0.36$& $35.4$\\
$\Ng \ge 13$  &$18.14\pm0.43$&$2.24\pm0.15$&$0.33$& $40.0$\\
$\Ng \ge 16$  &$19.33\pm0.48$&$2.56\pm0.23$&$0.48$& $46.9$\\
\hline
\hline
\multicolumn{5}{l}{$\sigma_z=0.007$}\\
\hline
$\Ng \ge 10$  &$14.80\pm0.46$&$1.98\pm0.10$&$0.42$& $33.1$\\
$\Ng \ge 11$  &$15.27\pm0.52$&$1.95\pm0.11$&$0.35$& $35.4$\\
$\Ng \ge 13$  &$17.14\pm0.56$&$2.00\pm0.13$&$0.31$& $40.0$\\
$\Ng \ge 16$  &$19.18\pm0.58$&$2.26\pm0.17$&$0.47$& $46.9$\\
\hline
\hline
\multicolumn{5}{l}{$\sigma_z=0$}\\
\hline
$\Ng \ge 10$  &$14.42\pm0.73$&$1.84\pm0.11$&$0.43$& $33.1$\\
$\Ng \ge 11$  &$14.81\pm0.81$&$1.80\pm0.12$&$0.39$& $35.4$\\
$\Ng \ge 13$  &$16.94\pm0.92$&$1.85\pm0.14$&$0.30$& $40.0$\\
$\Ng \ge 16$  &$20.10\pm0.92$&$2.13\pm0.19$&$0.49$& $46.9$\\
\end{tabular}
\end{table}
\clearpage

\begin{figure}[t]
\begin{center}
\includegraphics[width=.95\columnwidth]{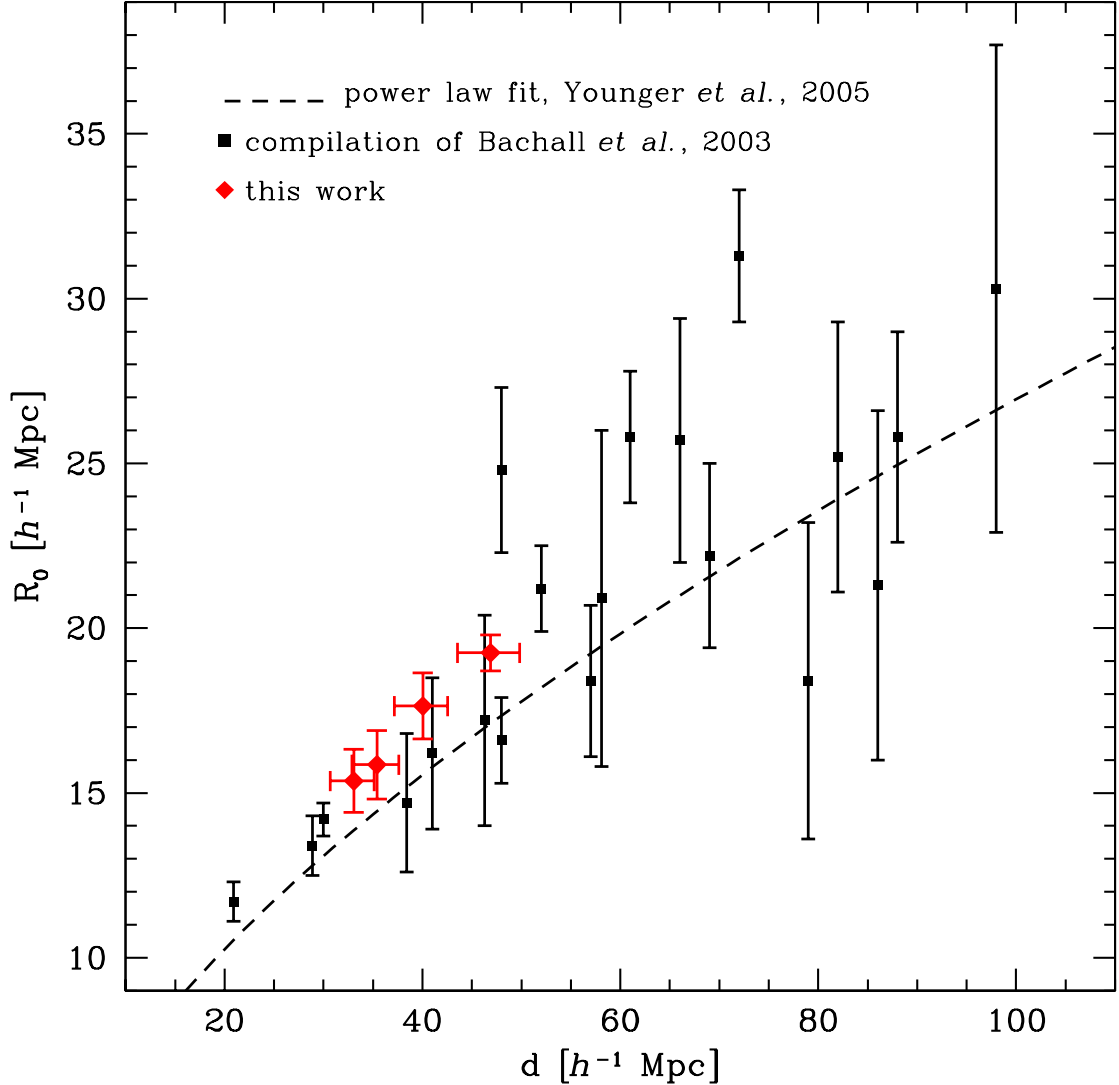}
\caption{\label{fig:r0d} 
Correlation length $R_0$ vs. mean cluster separation $d$ for different cluster 
samples. Results for the MaxBCG catalog (this work) are shown as red diamonds. 
Vertical error bars include the systematic uncertainty in $R_0$ due to photo-$z$ 
error uncertainty as well as the marginalized error from the $\chi^2$ 
analysis; horizontal error bars correspond to an assumed 20\% systematic 
uncertainty in the maxBCG sample volume. Black squares are from the 
compilation of earlier cluster measurements of \cite{bahcall_data}, in 
which a power-law slope $\gamma=2$ has been assumed. The dashed 
curve is the power-law fit of Eqn.~(\ref{eq:r0sim}) to the prediction of 
$\Lambda$CDM, from \cite{bahcall_sim}.}
\end{center}
\end{figure}
\clearpage

Models of structure formation predict that the correlation length $R_0$ should 
scale with the cluster mean separation $d$, where $1/d^3 = n_c$  
\citep{cc-scale_1,cc-scale_2,cc-scale_3,cc-scale_4}. 
\cite{bahcall_sim} studied this scaling using N-body 
simulations of the $\Lambda$CDM model and found that the results are well 
represented by a power-law relation over the separation range 
$20 \Mpc < d < 60 \Mpc$, 
\beq\label{eq:r0sim}
R_0 = 1.7 \left(\frac{d}{\Mpc}\right)^{0.6}\Mpc\,, 
\eeq
for $\sigma_8=0.84$. In Fig.~\ref{fig:r0d} we plot $R_0$ vs. $d$ for the MaxBCG samples (red diamonds) and for the cluster measurement compilation presented in 
\cite{bahcall_data} (black squares) and compare those  with the power-law 
$\Lambda$CDM  relation of Eqn.~(\ref{eq:r0sim}) (dashed curve). We have not plotted the SDSS EDR results of \cite{bahcall_data}, since they come from a subset of the current data and are therefore not independent of the new results we show here. 

As Fig.~\ref{fig:r0d} shows, the correlation function results given here are broadly consistent with those from previous cluster samples, including the scaling of the correlation length with richness or mean separation. However, the correlation amplitude is slightly higher than the $\Lambda$CDM prediction of Eqn.~(\ref{eq:r0sim}), by about 14 (8) \% for $\sigma_z=0.01$ (0.007); the cluster bias is higher by the same factor. To decide whether this difference is significant will require more precise modeling of the maxBCG photo-$z$ error distribution as a function of richness and redshift. In addition, Eqn.~(\ref{eq:r0sim}) has been derived for mass-selected catalogs, and it therefore uses a different selection function from the observations.

As noted in \S \ref{subsubsec:compare}, 
we can compare the results for a photo-$z$ dispersion of $\sigma_z=0.01$ with those for $\sigma_z=0.007$ to estimate the systematic error due to the uncertainty on the dispersion $\sigma_z$. For the $\Ng \ge 10$ sample, the resulting systematic error on the correlation length is 
$\Delta R_0 = 1.13$, about three times larger than the statistical error; for the correlation 
slope the systematic error is $\Delta \gamma =0.22$, about twice the statistic error. 
Including this systematic largely eliminates the discrepancy between the data and the model of 
Eqn.~(\ref{eq:r0sim}), as Fig. \ref{fig:r0d} shows. As noted in \S \ref{subsubsec:compare}, 
this systematic can be reduced by more careful modeling of $\sigma_z$ and its 
redshift dependence.

\subsection{Correlation function on large scales}
\label{subsec:analyse}

Here we consider the cluster correlation function measured over a larger range of scales, from $20$ to $195\Mpc$, and compare the results with the theoretical model of \S \ref{sec:model}. We describe 
the model correlation function for each MaxBCG sample by a two-parameter 
model, 
\beq
\tilde{\xi}_{cc}(\tr;s,b)=b^2~\tilde{\xi}_{mm}(\tr*s)\,,
\eeq
where $\tr$ is the separation in photo-$z$ space, $\tilde{\xi}_{mm}$ is the matter correlation function of Eqn. (\ref{eq:xil}), 
corrected for photometric redshift errors as described by Eqn.~(\ref{eq:xi_ph}), assuming
$\sigma_z=0.01$, and the parameter $s$ is described below. 
For comparison, we also show results for the model fits using the 
analytic correction of Eqn.~(\ref{eq:ps_iso}) for the photo-$z$ errors. Note that, 
in fitting for the cluster linear bias parameter $b$, we fix the linear matter power 
spectrum amplitude $\sigma_8$ to its fiducial value of 0.9; in actuality, the fit  
constrains the product $b(\sigma_8/0.9)$. 

In this model, we have assumed that small variations in the  cosmological parameters 
can be incorporated simply as changes in the predicted physical separation $r$, {\it i.e.}, 
by the scale shift parameter $s$ defined by 
\beq
s\equiv \left[\frac{D_M^2(z;p)}{H(z;p)}\right]^{1/3}
\left[\frac{H(z;p^*)}{D_M^2(z;p^*)}\right]^{1/3}\,.
\eeq
Here, $z$ is the median survey redshift, 
$H(z;p)$ is the Hubble parameter, $D_M(z;p)=[c/(1+z)]\int_0^z dz'/H(z')$ is the comoving angular diameter distance, and $p$ represents the cosmological parameters 
\citep{Seo2003,Eisenstein2005,Angulo2007,Seo2007,Huetsi2007}; $p*$ represents the fiducial set of cosmological parameters enumerated at the end of \S \ref{subsec:est}. The scale 
shift parameter involves a geometric average of the two transverse components and the line-of-sight component and therefore applies to spherically averaged separations $r$. Although the 
$s$ parameter does {\it not} capture the full cosmological parameter dependence of the 
correlation function, it does describe the effects of small cosmological parameter variation 
on the location of the baryon acoustic peak to the accuracy we need \citep{blakeglazebrook}.
It is therefore a convenient model parametrization for the purpose of 
determining the significance of the BAO feature. 

To assess the significance of a possible BAO feature, we also compare the data to a model with no BAO feature in the linear power spectrum, computed using the smooth transfer function of \cite{EisensteinHu1998}. In the no-BAO case, the uncertainties in the measurements of the cluster correlation functions do not allow a meaningful constraint on the shift parameter $s$: with a flat prior on it, $s$ tends to unphysically small values, particularly if we restrict the analysis to relatively large scales ($r\gtrsim 60\Mpc$). In the computation of the $\chi^2$ statistic, we therefore introduce, for {\it both} the BAO and no-BAO cases, a Gaussian prior on the parameter $s$, with central value $s=1$ (corresponding to our fiducial cosmology) and standard deviation of $0.05$, consistent with the current uncertainties in the relevant cosmological parameters $\Omega_m h^2$ and $\Omega_b h^2$ from WMAP CMB observations \citep{WMAP3}. 
Since we are not attempting to constrain cosmological parameters with this 
measurement but only gauging the significance of a possible BAO feature, 
this prior on $s$ simply allows us to make a sensible comparison between the
BAO and no-BAO models.

\clearpage
\begin{figure*}[p]
\begin{center}
\includegraphics[width=0.98\textwidth]{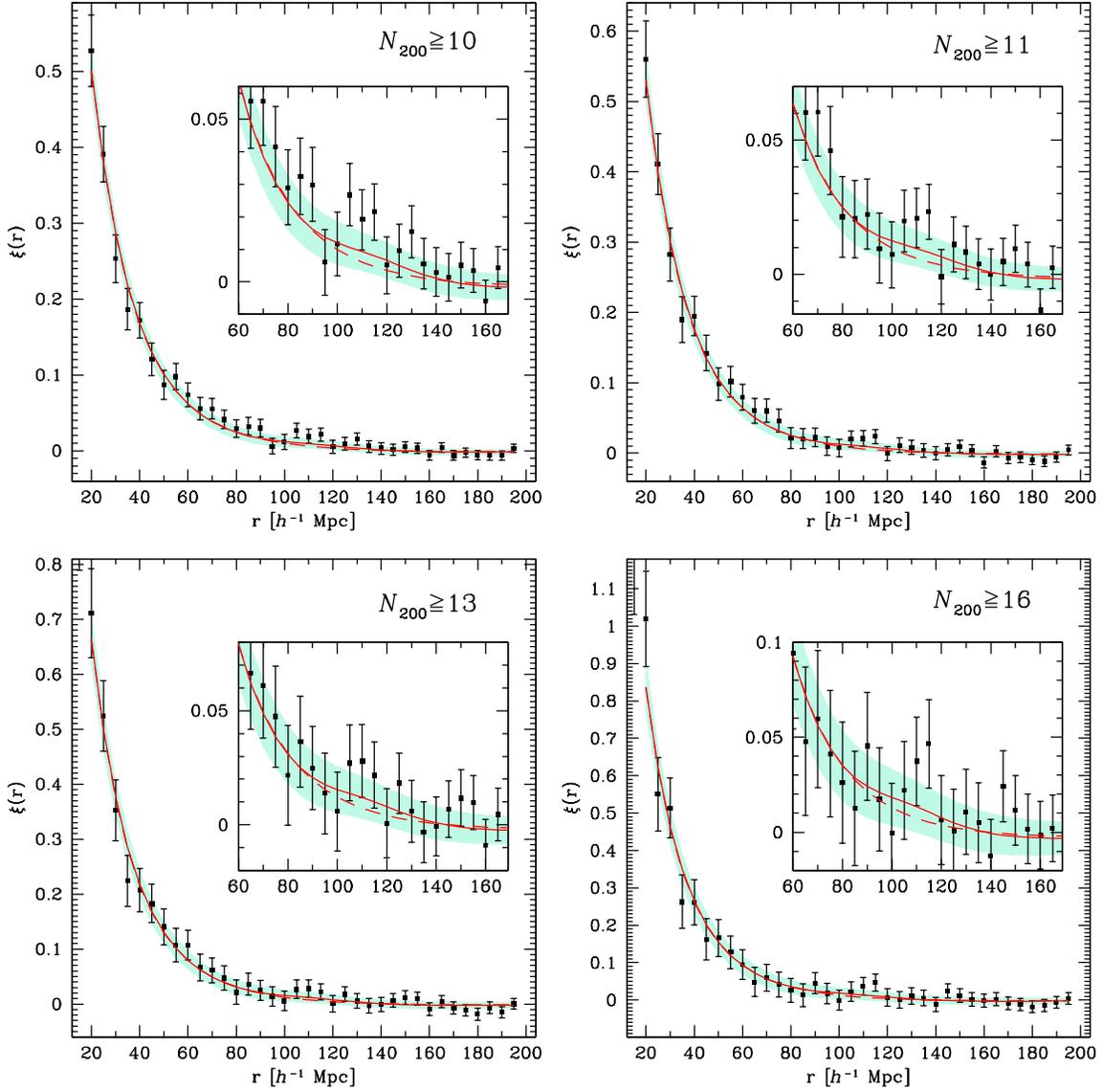}
\caption{\label{fig:cf_results} Measured correlation functions for the four MaxBCG samples (points) in 36 linear separation bins from $r=20$ to $195 \Mpc$. 
Best-fit $\Lambda$CDM models from the two-parameter fits with BAO features (solid red curves) and without acoustic features (dashed red curves). Error bars on the data points are estimated using the jackknife, while the shaded green areas show the linear theory (Gaussian) predictions for the errors.
Insets show close-ups of the region around the expected BAO feature.
}
\end{center}
\end{figure*}
\begin{figure*}[!hp]
\begin{center}
\includegraphics[width=0.98\textwidth]{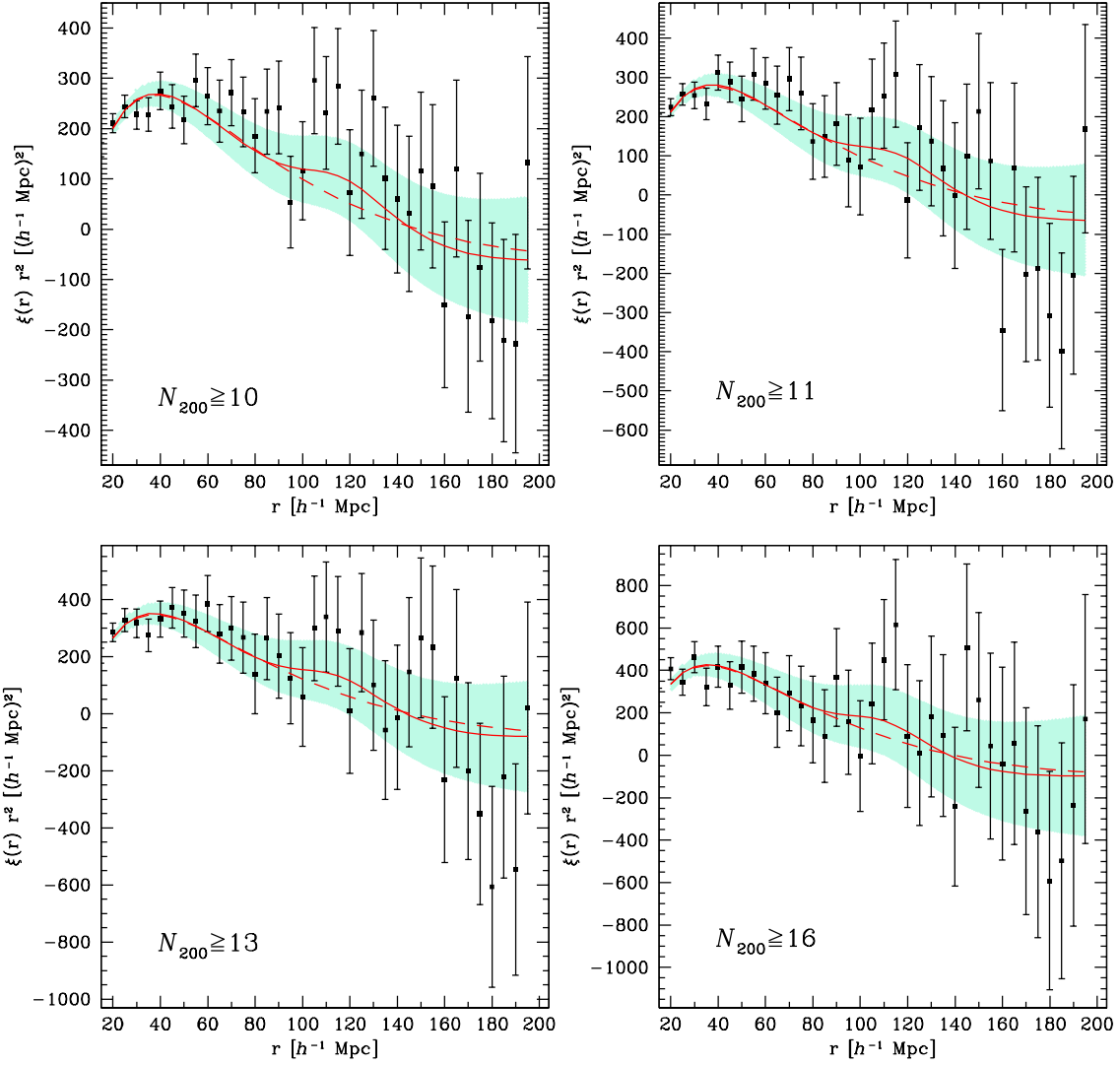}
\caption{\label{fig:cf_results_cfr2} Same as fig.~\ref{fig:cf_results} but 
showing  $\xi(r) r^2$ to emphasize the large-scale features.}
\end{center}
\end{figure*}
\clearpage

\subsubsection{Results}

Figs.~\ref{fig:cf_results} and \ref{fig:cf_results_cfr2} show 
the measured cluster correlation functions together with the best-fit models from the two-parameter analysis both with and without acoustic features and using the geometric correction for photo-$z$ errors. The error bars on the data points correspond to one standard deviation from the diagonal of the jackknife covariance matrix; since the covariance matrix is non-diagonal, the points in different separation bins are correlated. For each sample, the green shaded region shows the Gaussian prediction for the standard deviation, plotted around the best-fit model.

In Table~\ref{tab:results} we present the best-fit values for the two parameters $b$ and $s$ for the four MaxBCG samples and the corresponding $\chi^2$ values. The fits are based on measurements in 36 linear bins ranging in separation $\tr$ from $20$ to $195\Mpc$, resulting in $34$ degrees of freedom. For the upper rows of the Table, the correlation function covariance is determined from the jackknife method; the lower rows show results using the linear perturbation theory covariance matrix, Eqn.~(\ref{eq:covar_linear}). Note that the linear theory covariance estimate requires knowledge of the linear bias factor $b$, which is one of the parameters we are aiming to extract from the data. We therefore first estimate the linear covariance using the values of $b$ from the jackknife error fits; the resulting linear theory covariance estimates 
are then used to recompute the $\chi^2$ values and extract new estimates of $b$ and $s$.   

\clearpage
\begin{table}[b]
\caption{\label{tab:results} Best-fit values for the parameters $b$, $s$ from $\tilde{\xi}_{cc}
(\tr)$ measurements for the four MaxBCG samples, using the covariance from the jackknife method (upper rows) and from the Gaussian analytic prediction (lower rows), with the error on $b$ marginalized over $s$. Last column displays $\chi^2$ difference between the best BAO and no-BAO fits. The fits use the geometric correction for photo-$z$ errors with $\sigma_z=0.01$ and measurements over the range $20\le \tr \le 195\Mpc$. The $\chi^2$ value corresponds to $34$ degrees of freedom.} 
\begin{tabular}{c|ccc|ccc|c}
 sample&\multicolumn{3}{c}{BAO fits}&\multicolumn{3}{c}{no-BAO fits}&\\ 
 $\Ng\ge$       &$b$&$s$&$\chi^2$&$b$&$s$&$\chi^2$&$\Delta\chi^2$\\
\hline
\multicolumn{8}{l}{Covariance matrix from jackknife}                \\
\hline
$10$    &$2.80\pm 0.13$&$0.96$&$24.6$&$2.79\pm 0.14$&$0.96$&$26.6$&$2.0$\\
$11$    &$2.91\pm 0.15$&$0.97$&$22.3$&$2.90\pm 0.16$&$0.98$&$23.8$&$1.5$\\
$13$    &$3.26\pm 0.20$&$0.97$&$18.0$&$3.25\pm 0.20$&$0.98$&$18.9$&$0.9$\\
$16$    &$3.76\pm 0.24$&$1.02$&$19.2$&$3.74\pm 0.25$&$1.03$&$20.5$&$1.3$\\
\hline
\multicolumn{8}{l}{Covariance matrix from linear theory}             \\
\hline
$10$    &$2.86\pm 0.11$&$0.98$&$93.1$&$2.86\pm 0.13$&$1.00$&$ 95.5$&$2.4$\\
$11$    &$2.95\pm 0.12$&$0.97$&$97.9$&$2.95\pm 0.13$&$0.99$&$100.2$&$2.3$\\
$13$    &$3.35\pm 0.14$&$1.00$&$77.9$&$3.37\pm 0.16$&$1.02$&$ 79.7$&$1.8$\\
$16$    &$3.96\pm 0.19$&$1.04$&$76.5$&$3.98\pm 0.20$&$1.07$&$ 78.4$&$1.9$\\
\end{tabular}
\end{table}
\clearpage

Using the jackknife error estimates, the $\Ng \ge 10$ sample shows a difference in $\chi^2$ between the BAO and no-BAO models of $\Delta \chi^2 =2$; this corresponds to a marginal significance of $1.4\sigma$ for the BAO feature. 
Using the same sample, \cite{Huetsi2007} found a somewhat larger significance of about $2\sigma$ for the BAO feature from a power spectrum analysis. Using the jackknife error and computing the {\it theoretically expected} $\chi^2$ difference for this sample between the BAO and no-BAO models, the typical expected difference corresponds to only $1\sigma$. That is, given the photo-$z$ errors and the sample size, we would not expect {\it a priori} to find a highly significant BAO detection from this sample, as noted in \S \ref{subsubsec:compare}. The significance of the BAO feature generally goes down as the cluster richness threshold is increased, reflecting the larger Poisson errors for these smaller samples. The values for the bias parameter trend upward with increasing cluster richness, as  expected on theoretical grounds---more massive clusters are more strongly clustered (see \S \ref{subsec:clusterbias})---and consistent with the results on smaller scales (\S \ref{subsec:estimate}).

If we drop the 5\% Gaussian prior on the shift parameter $s$, the correlation function data of the $\Ng\ge 10$ sample constrains it to the $1\sigma$ range $s = 0.92 \pm 0.08$ for the BAO model fit and the jackknife  
covariance matrix. The $8\%$ error on $s$, roughly twice that for the SDSS spectroscopic LRG sample \citep{Eisenstein2005}, does not yield a significant 
constraint on cosmological parameters. 

The same analysis carried out using the Gaussian (linear theory) error covariance yields significantly larger values for the $\chi^2$ statistic in all samples, a sign that this method possibly underestimates 
the errors. However, the  differences in $\chi^2$ between the BAO and no-BAO models are comparable to 
though slightly larger than those for the jackknife errors. In the $\Ng\ge 10$ case, for instance, $\Delta\chi^2=2.4$, corresponding to a significance of $1.5\sigma$.

\subsubsection{Analytic photo-z error correction}

\clearpage
\begin{table}[b]
\caption{\label{tab:results_pzA} Same as Table~\ref{tab:results} but using the analytic photo-$z$ error correction of Eqn.~(\ref{eq:ps_iso}).} 
\begin{tabular}{l|ccc|ccc|c}
 sample&\multicolumn{3}{c}{BAO fits}&\multicolumn{3}{c}{no-BAO fits}&\\ 
$\Ng\ge$   &$b$&$s$&$\chi^2$&$b$&$s$&$\chi^2$&$\Delta\chi^2$\\
\hline
\multicolumn{8}{l}{Covariance matrix from jackknife}                \\
\hline
$10$    &$2.84\pm 0.14$&$0.94$&$24.3$&$2.82\pm 0.15$&$0.95$&$26.7$&$2.4$\\
$11$    &$2.96\pm 0.16$&$0.96$&$21.9$&$2.94\pm 0.17$&$0.97$&$23.6$&$1.7$\\
$13$    &$3.32\pm 0.20$&$0.97$&$17.6$&$3.30\pm 0.21$&$0.98$&$18.7$&$1.1$\\
$16$    &$3.82\pm 0.25$&$1.01$&$19.0$&$3.82\pm 0.27$&$1.02$&$20.5$&$1.5$\\
\hline
\multicolumn{8}{l}{Covariance matrix from linear theory}             \\
\hline
$10$    &$2.89\pm 0.12$&$0.96$&$91.7$&$2.88\pm 0.14$&$0.98$&$ 94.7$&$3.0$\\
$11$    &$2.98\pm 0.13$&$0.96$&$96.4$&$2.98\pm 0.14$&$0.98$&$ 99.2$&$2.8$\\
$13$    &$3.38\pm 0.15$&$0.98$&$76.7$&$3.38\pm 0.17$&$1.00$&$ 79.0$&$2.3$\\
$16$    &$3.99\pm 0.20$&$1.02$&$75.6$&$4.00\pm 0.22$&$1.04$&$ 78.1$&$2.5$\\
\end{tabular}
\end{table}
\clearpage

\begin{figure}[tb]
\begin{center}
\includegraphics[width=.95\columnwidth]{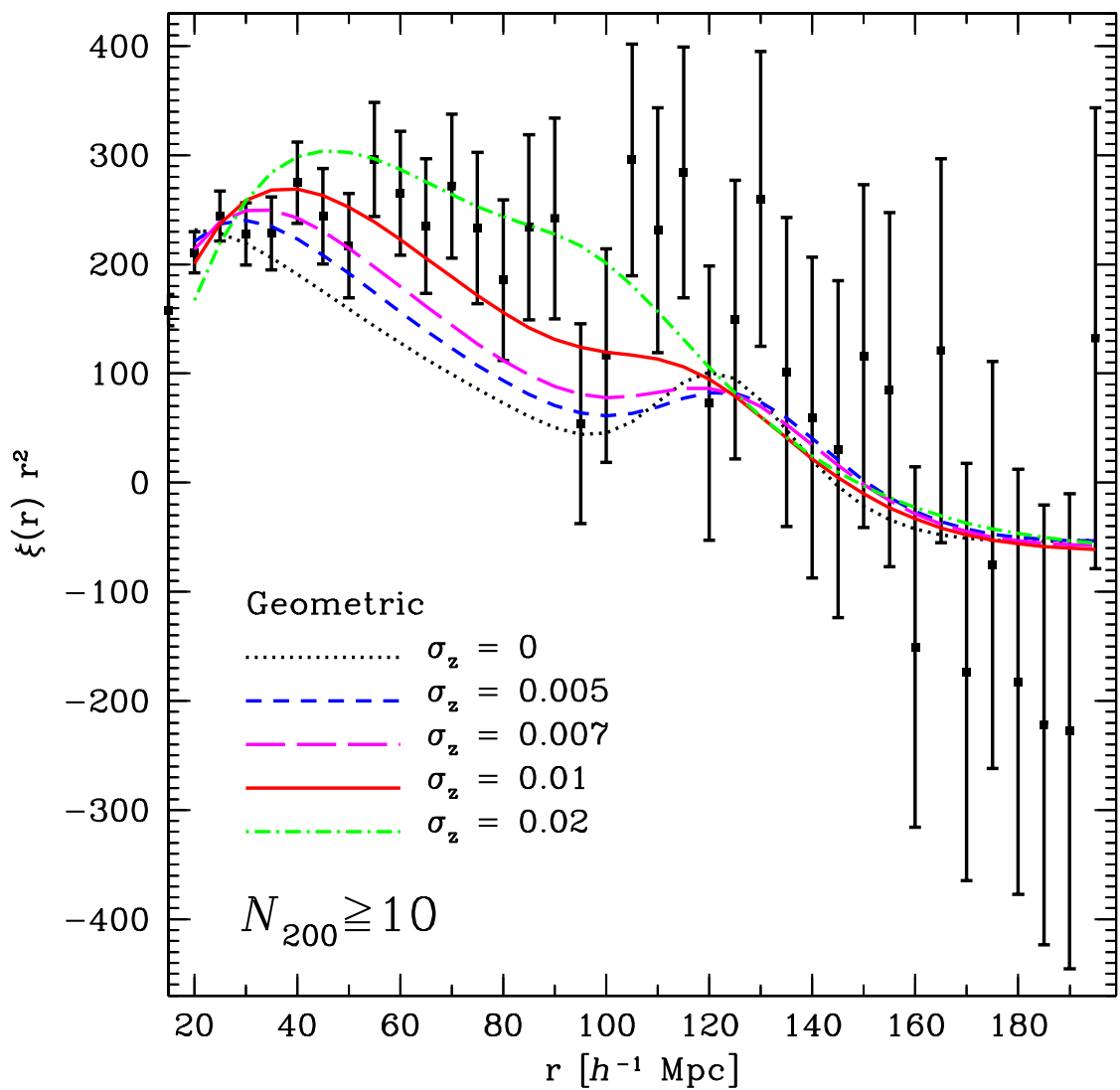}
\caption{\label{fig:cf_N10_bf_different_dz} Best-fit models to $r^2 \xi(r)$ for the $\Ng\ge 10$ sample, assuming different errors on the photometric redshift determination: $\sigma_z=0$ (no correction, dotted black curve), $\sigma_z=0.005$ (short-dashed blue curve), $\sigma_z=0.007$ (long-dashed magenta curve), $\sigma_z=0.01$ (continuous red curve), and $\sigma_z=0.02$ (dot-dashed green curve), using the geometric 
photo-$z$ correction. }
\end{center}
\end{figure}
\clearpage

To gauge the robustness of these results, in Table~\ref{tab:results_pzA} we present the model fits using the analytic photo-$z$ correction of Eqn.~(\ref{eq:ps_iso}) in place of the geometric correction. The best-fit values of the parameters as well as the absolute values and differences in $\chi^2$ are close to and consistent with those presented above. The values of the bias parameters and the significance of the BAO feature are both slightly higher in this case. This is traceable to the fact that the analytic photo-$z$ correction leads to slightly less smoothing of the correlation function than the geometric correction for $\sigma_z=0.01$, as can be seen from Fig.~\ref{fig_cf_photoz}. On one hand, reduced smoothing decreases the correlation amplitude at small scales, where the error bars are smaller. On the other hand, a more pronounced (less smoothed) acoustic peak better fits the excess of power at scales of about $110\Mpc$ that is clearly visible in the data in Fig.~\ref{fig:cf_results_cfr2}. 
For the $\Ng\ge 10$ sample, the significance of the BAO feature here is 
about $1.5$ - $1.7\sigma$, depending on the covariance estimate (jackknife or 
linear), closer to the result of \citep{Huetsi2007}, which used the same analytic photo-$z$ 
error correction to the power spectrum.

\subsubsection{Dependence on photometric redshift error}
\label{subsubsec:photozresults}

\clearpage
\begin{table}[b]
\caption{\label{tab:results_dz} Best-fit values for the 2-parameter analysis
of the $\Ng\ge 10$ sample assuming different values for the photometric redshift 
error $\sigma_z$. The middle rows correspond to the geometric correction, the lower ones to the analytic method. Here we assume jackknife covariance, and there are 
$34$ d.o.f. for the $\chi^2$ analysis.}
\begin{tabular}{l|ccc|ccc|c}
      &\multicolumn{3}{c}{BAO fits}            &\multicolumn{3}{c}{no-BAO fits}  &   \\
$\sigma_z$& $b$ & $s$  &$\chi^2$& $b$  & $s$   &$\chi^2$& $\Delta\chi^2$ \\
\hline
\multicolumn{8}{l}{No photo-$z$ correction}                               \\
\hline
$0$       &$2.30$&$0.90$&  $39.6$    &$2.17$ &$0.86$ &$43.6$ & $4.0$  \\
\hline
\multicolumn{8}{l}{Geometric photo-$z$ correction}           \\
\hline
$0.005$   &$2.27$&$0.88$&  $31.8$    &$2.19$ &$0.86$ &$33.8$ & $2.0$  \\
$0.007$   &$2.44$&$0.90$&  $27.8$    &$2.39$ &$0.89$ &$29.6$ & $1.8$  \\
$0.01$    &$2.81$&$0.96$&  $24.6$    &$2.79$ &$0.96$ &$26.6$ & $2.0$  \\
$0.02$    &$4.20$&$1.14$&  $36.1$    &$4.19$ &$1.15$ &$36.2$ & $0.1$  \\
\hline
\multicolumn{8}{l}{Analytic photo-$z$ correction}\\
\hline
$0.005$   &$2.28$&$0.88$&  $31.7$    &$2.22$ &$0.86$ &$33.2$ & $1.5$  \\
$0.007$   &$2.51$&$0.91$&  $27.6$    &$2.45$ &$0.90$ &$29.5$ & $1.9$  \\
$0.01$    &$2.84$&$0.94$&  $24.3$    &$2.82$ &$0.95$ &$26.7$ & $2.4$  \\
$0.02$    &$3.85$&$1.02$&  $25.0$    &$3.88$ &$1.04$ &$26.3$ & $1.3$  \\
\end{tabular}
\end{table}
\clearpage

Here we assess the impact of uncertainties in the photometric redshift errors on the results. We carry out the analysis of the $\Ng\ge 10$ sample using different values of the photo-$z$ dispersion $\sigma_z$. In Table~\ref{tab:results_dz} we present the best-fit parameter and $\chi^2$ values for $\sigma_z=0$, $0.005$, $0.01$, and $0.02$ for both the geometric and analytic photo-$z$ error corrections. The significance of the BAO feature ($\Delta \chi^2$) is not strongly dependent on the photo-$z$ dispersion, except for the case $\sigma_z=0.02$, which is twice as large as estimated from the maxBCG catalog. Fig.~\ref{fig:cf_N10_bf_different_dz} shows the best-fit models of table~\ref{tab:results_dz} using the geometric photo-$z$ correction, compared to the data for the $\Ng\ge 10$ sample. 

As we discussed in \S \ref{subsec:estimate}, we can use the difference in 
results for $\sigma_z=0.01$ and $\sigma_z=0.007$ to derive an approximate systematic error estimate for the bias parameter due to the uncertainty in $\sigma_z$. For the $\Ng\ge 10$ sample, a difference of 
$\Delta \sigma_z = 0.003$ corresponds to a difference of $\Delta b = 0.37$. As we noted 
for the correlation length in \S \ref{subsec:estimate}, this error is significantly larger than the 
statistical error of about $0.14$ for this sample. However, this systematic error 
estimate is quite conservative. Moreover, any redshift dependence of $\sigma_z$ can be included in 
the geometric method and should not be counted as a systematic error. We postpone a more detailed 
discussion of this point to a future paper.

\subsubsection{The integral constraint}

So far we have not modeled the impact of the integral constraint, which arises from the fact that the integral of the two-point function over the survey volume is assumed to be zero when estimating the sample density in Eqn. (\ref{eq:LS}). One can model this by including an additive constant $c$ in the correlation function,  
\beq
\tilde{\xi}_{obs}=b^2\left[\tilde{\xi}_{mm}(\tr;s)+c\right] ~.
\eeq
We find that the improvement to the model $\chi^2$ by treating this term as an extra free parameter is very small. For instance, for the $\Ng\ge 10$ sample, the analysis of this three-parameter model gives a marginalized value of $c=0.0002\pm 0.0003$, consistent with zero, while it changes the best-fit values for $b$ and $s$ by amounts much smaller than their statistical errors. Compared to the two-parameter model above, the value of the best-fit $\chi^2$ decreases by only $0.4$ (for 34 d.o.f.) for both the BAO and no-BAO models. 

\clearpage
\begin{figure}[t]
\begin{center}
\includegraphics[width=0.98\columnwidth]{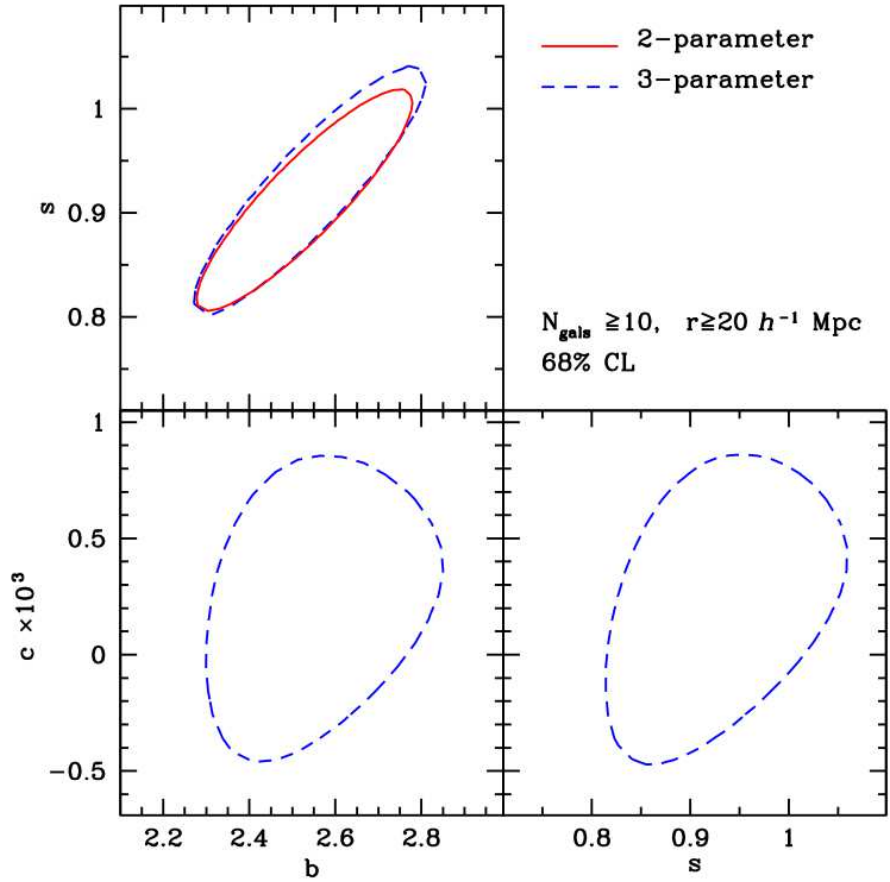}
\caption{\label{fig:contours} 1-$\sigma$ contour plots for $b$, $s$, and $c \times 10^{3}$ 
(dashed, blue curves) marginalized over the third parameter, for the $\Ng\ge 10$ sample..  
Upper left panel also shows results of a 2-parameter 
analysis ($b$ and $s$: red, continuous line) for comparison.  
}
\end{center}
\end{figure}
\clearpage

To illustrate these points, in the upper left panel of Fig.~\ref{fig:contours} we show the 1-$\sigma$ contour for the 2-parameter analysis (red, continuous line) compared to the 3-parameter analysis in which we marginalize over $c$ (blue, dashed lines) for the $\Ng\ge 10$ sample. In the latter case, we do not impose the WMAP prior on the shift parameter $s$, and we only consider the baryonic model. Other panels show contours for $c$ versus $b$ and $s$, marginalized over the missing parameter. The degeneracy between the bias and shift parameters is not significantly affected by the introduction of $c$. The additive constant also does not introduce large degeneracies with the other two parameters, and its allowed values are consistent with zero.

\subsubsection{Sample Purity}

According to \cite{maxBCGsample}, the MaxBCG cluster sample is estimated to be about 90\% pure. 
As a simple test of the effects of sample purity on the correlation function, we replace $10\%$ of the clusters in the $\Ng\ge 10$ sample with randomly distributed points. As expected, this lowers the correlation function amplitude by a factor of about $(0.9)^2$, corresponding to a $\sim 10$\% systematic uncertainty in the 
bias, with negligible change in the shift parameter $s$. We also find slightly lower values for the $\chi^2$, particularly for the no-BAO models, when the jackknife covariance and the geometric correction for photo-$z$ 
errors are used. This results in a significantly lower value for the $\chi^2$ difference 
between the BAO and the no-BAO model, $\Delta \chi^2=0.7$.

\subsection{Cluster bias}
\label{subsec:clusterbias}

Here we compare the values obtained for the cluster bias parameters for the different richness samples with the theoretical predictions that can be derived in the framework of the Halo Model  \citep{MoWhite1996,MoJingWhite1996,ShethTormen1999}. We also translate the bias-richness relation into a measurement of bias vs. halo mass, using the mass-richness relation derived from statistical weak lensing measurements for this cluster sample in \citep{Johnston2007}.

\begin{figure}[t]
\begin{center}
\includegraphics[width=0.98\columnwidth]{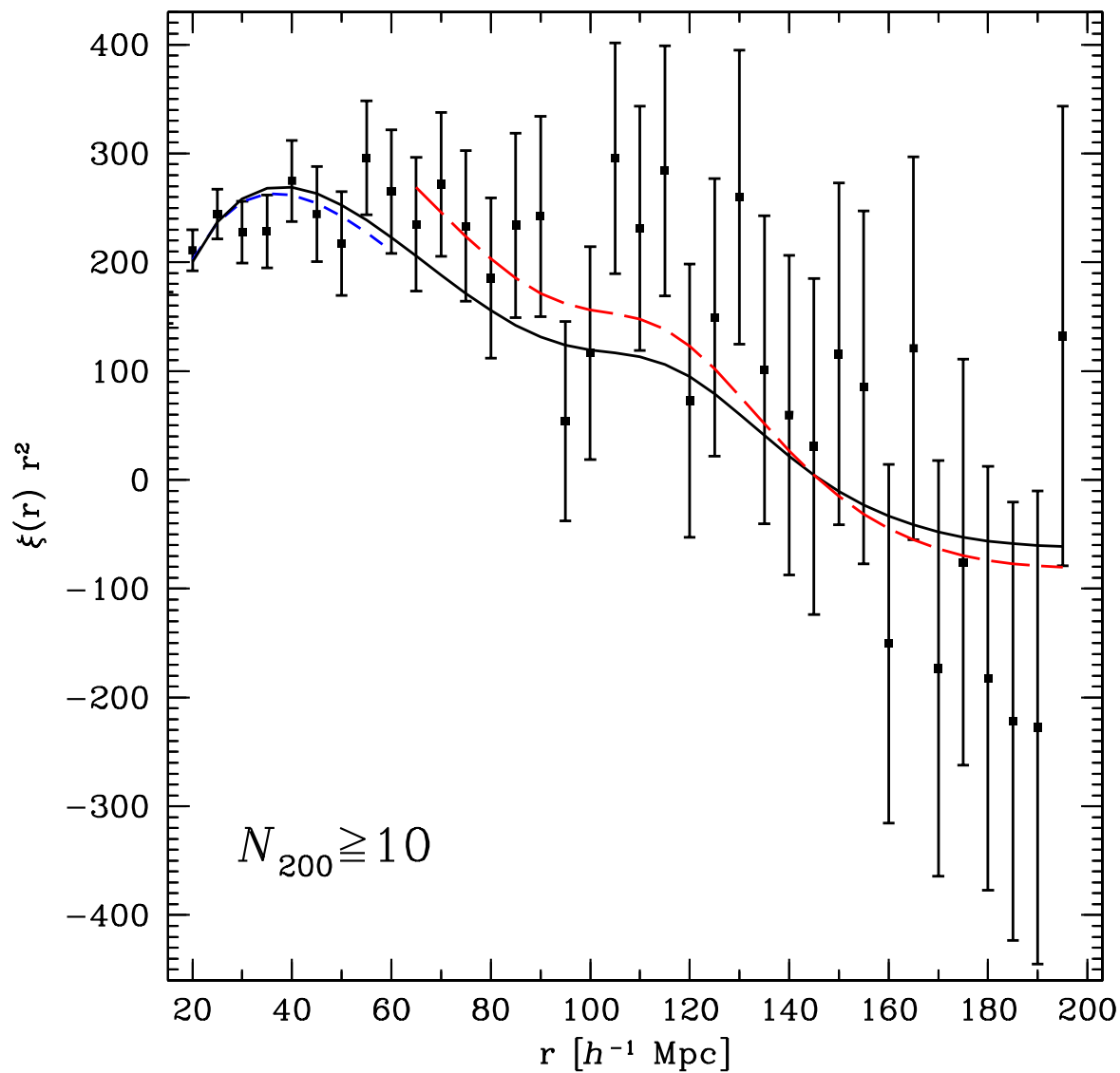}
\caption{\label{fig:bias_fits} 
Two-parameter $\Lambda$CDM fits to the correlation function of the $\Ng\ge 10$ sample, using different separation ranges: $r=20-60 \Mpc$ (short-dashed blue), $20-195 \Mpc$ (solid black), and $60-195 \Mpc$ (long-dashed red). Here we use jackknife error covariance and the geometric photo-$z$ error model with $\sigma_z=0.01$.}
\end{center}
\end{figure}

For this analysis, we compare results using three different separation intervals: $r=20-60 \Mpc$, $20-195 \Mpc$, and $60-195 \Mpc$. The resulting values for the linear bias are given in 
Table~\ref{tab:results_bias}, and the corresponding fits to the data of the $\Ng\ge 10$ sample are plotted in Fig.~\ref{fig:bias_fits}. As in the previous section, the $1-\sigma$ errors on $b$ are determined after marginalizing over the shift parameter $s$, including the $5\%$ Gaussian CMB prior on $s$. The $\Lambda$CDM model with BAO is assumed.

\clearpage
\begin{table}[b]
\caption{\label{tab:results_bias} 
Linear cluster bias parameter $b$ estimated for the different samples, using different ranges in the separation $\tr$ and the jackknife covariance matrix. The values have been marginalized over the shift parameter $s$, using a Gaussian prior with mean $s=1$ and dispersion $\sigma_s=0.05$. We use the geometric photo-$z$ error correction with $\sigma_z=0.01$ and assume $\sigma_8=0.9$ for the mass clustering amplitude.} 
\begin{tabular}{l|ccc}
 sample&\multicolumn{3}{c}{range~$[\Mpc]$}\\
$\Ng\ge$    &$20\le \tr \le 195$& $20\le \tr\le 60$ &$60\le \tr \le 195$\\
\hline
$10$    &$2.81\pm 0.13$&$2.87\pm 0.16$&$3.21\pm 0.37$\\
$11$    &$2.91\pm 0.15$&$2.96\pm 0.17$&$3.35\pm 0.44$\\
$13$    &$3.26\pm 0.19$&$3.33\pm 0.21$&$3.40\pm 0.55$\\
$16$    &$3.76\pm 0.24$&$3.82\pm 0.27$&$3.46\pm 0.77$\\
\end{tabular}
\end{table}
\clearpage

\clearpage

Table~\ref{tab:results_bias} indicates that the bias increases with richness, as noted above. The increase is driven by the measurements on scales $r \lesssim 60 \Mpc$: there is no significant trend of correlation amplitude with richness on larger scales, because the statistical errors are large there. In addition, for the $\Ng \ge 10$ and $11$ samples, the bias on scales $r \gtrsim 60 \Mpc$ appears to be larger than that on smaller scales, as indicated by the mismatch of the model fits in Fig.~\ref{fig:bias_fits}. This could be an indication of scale-dependent bias or of extra large-scale power beyond that expected in $\Lambda$CDM, as also suggested by the photometric LRG analyses \citep{Blake2007,Padmanabhan2007}, but the trend is not statistically significant given the errors for the current sample.  

\clearpage
\begin{figure*}[t]
\begin{center}
\includegraphics[width=0.4\columnwidth]{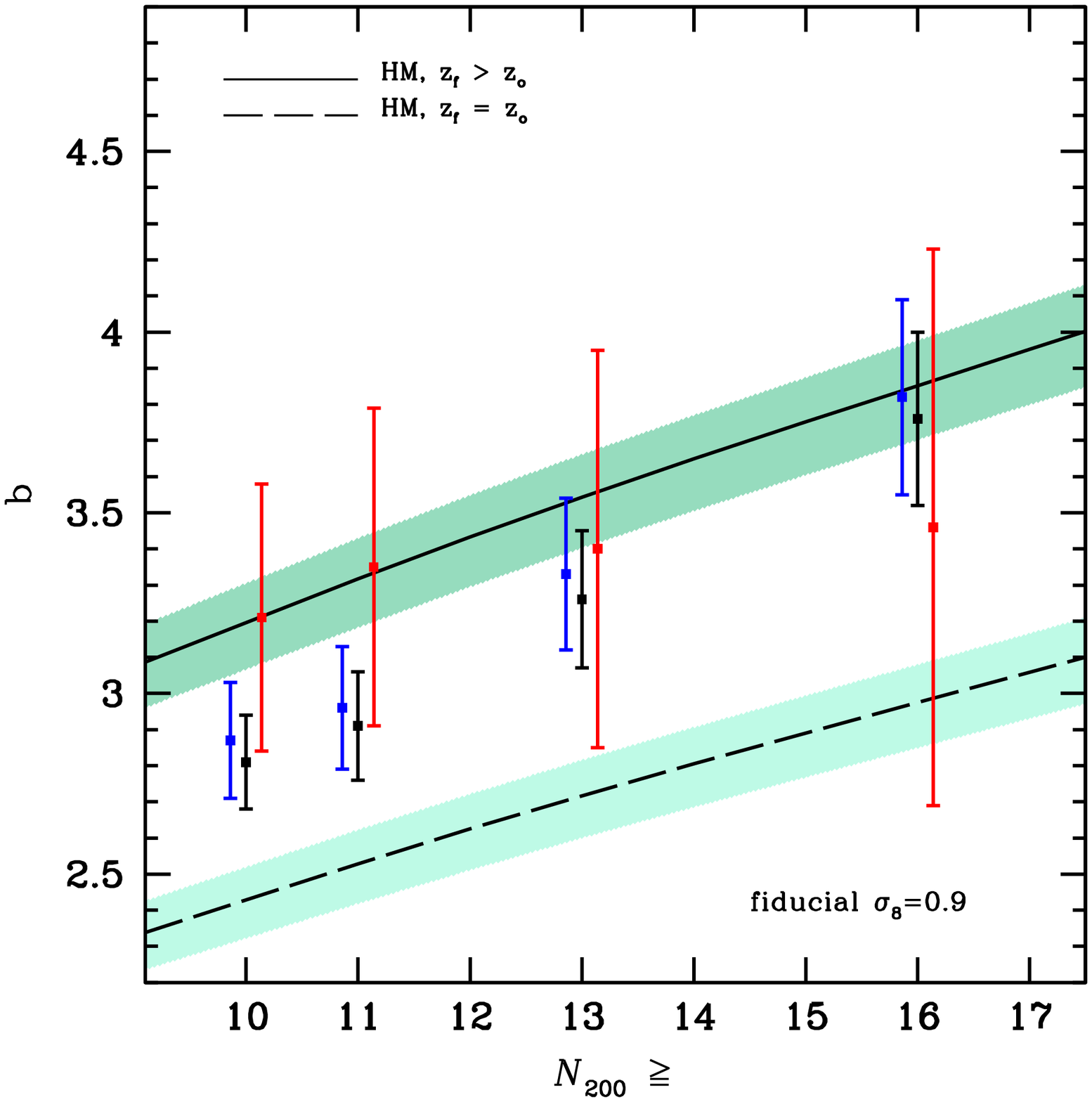}
\includegraphics[width=0.4\columnwidth]{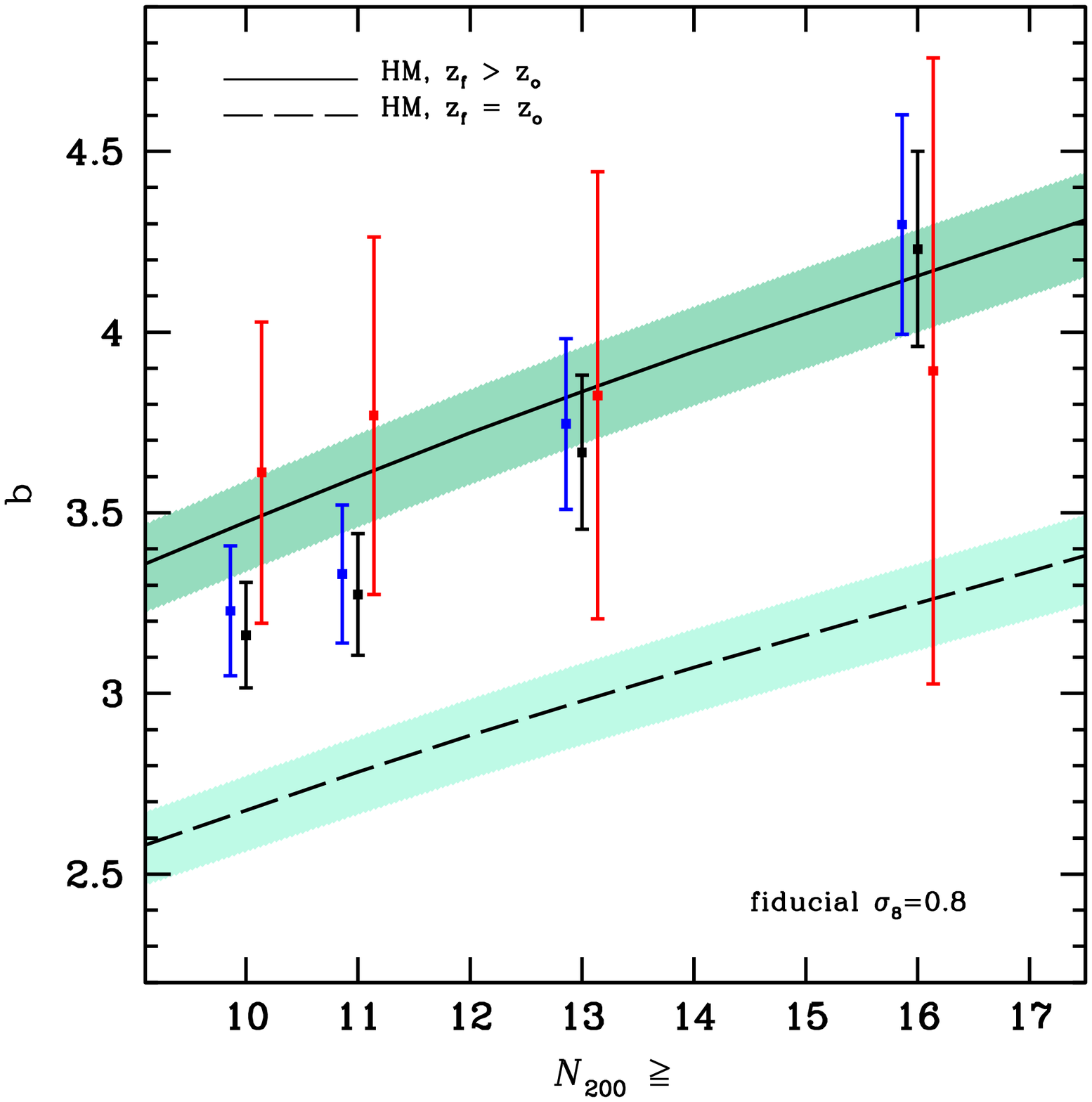}
\caption{\label{fig:bias} 
Cluster bias as a function of richness. The data points correspond to the values measured in the correlation function analysis, Table \ref{tab:results_bias}, with geometric correction for photo-$z$ errors. Points correspond to different separation ranges for the measurement: $r=20$ to $195\Mpc$ (black), $20-60 \Mpc$ (blue), and $60 - 195\Mpc$ (red); the points are slightly displaced from each other horizontally for clarity. The curves represent the Halo Model predictions, for which mass has been translated to richness by matching the cluster counts to the halo mass function at $z=0.22$. Dashed curve assumes $z_f=z_o$, and the continuous curve accounts for the distribution of formation redshifts [Eqn. 
(\ref{eq:halo_bias_zF})]. Shaded regions around the HM curves show the errors due to 20\% uncertainty in the survey volume. {\it Left panel:} assuming $\sigma_8=0.9$ for the linear power spectrum normalization; {\it Right panel:} $\sigma_8=0.8$. Here the bias values measured from the correlation function have been 
rescaled by $0.9/0.8$ since the clustering amplitude is proportional to 
$(b\sigma_8)^2$. }
\end{center}
\end{figure*}
\clearpage

The Halo Model (hereafter HM) provides an analytic expression for the bias of halos as a function of halo mass, $b_h(M)$ \citep{MoWhite1996,MoJingWhite1996,ShethTormen1999}. We can therefore express the expected value for the bias of a given cluster richness sample as
\beq\label{eq:bias}
b\simeq\frac{1}{n_c}\int_{M_{\smin}}^\infty \!\!\!dM n_h(M,z)~ b_h(M,z)\,,
\eeq
where $n_h(M,z)$ is the mass function of dark matter halos at redshift $z$, and $b_h(M,z)$ is the halo bias function from HM. The threshold mass $M_{min}$ for a given richness sample can be determined by requiring that the theoretical cluster number density,
\beq\label{eq:density}
n_c=\int_{M_{\smin}}^\infty \!\!\!dM n_h(M,z)\,,
\eeq
matches the observed value for the sample. The implicit simplifying assumption in this matching is that halo mass is a monotonic function of cluster richness, with no scatter in the relation between them. 

We use the Sheth \& Tormen \citep{ShethTormen1999} formula for $n_h(M,z)$, 
\beq
n_h(M,z)=-\frac{\bar{\rho}}{M^2}\frac{d\ln\sigma}{d\ln M}f(\nu)\,,
\eeq
where $\nu=\delta_c/\sigma(M,z)$ with $\delta_c=1.686$, $\sigma^2(M,z)$ is the variance in the linear density perturbation amplitude on mass scale $M$ at redshift $z$, and the function 
\beq
f(\nu)=A\sqrt{\frac{2q}{\pi}}\left[1+(q\nu^2)^{-p}\right]\nu e^{-q\nu^2/2}~,
\eeq
with $A=0.322$, $p=0.3$, and $q=0.707$. We neglect the mild dependence of the critical density $\d_c$ on the value of $\Omega_m(z)$, using the constant value from spherical collapse in an Einstein-de Sitter Universe. For our fiducial cosmology, and assuming the survey volume to be $0.5\cGpc$, by matching the number densities we find for our four samples the theoretical mass thresholds $M_{\smin}=8.6$, $9.9$, $12.5$ and $16.7\times 10^{13}\Ms$, respectively for $\Ng\ge10$, $11$, $13$ and $16$. Note that these mass estimates are about a factor of two higher than those inferred directly from weak lensing measurements for the same cluster sample (\S \ref{subsec:cat}) \citep{Johnston2007}, a point to which we return below.

The halo bias function $b_h$, which depends on both the redshift of observation $z_o$ and on the redshift of formation $z_f$ of the halos, is given in the HM by (see, e.g., \cite{MoJingWhite1996,ScoccimarroEtal2001})
\bea\label{eq:halo_bias}
b_h(M,z_o,z_f) & = & 1 + \frac{q\nu^2(z_f)-1}{\d_f(z_o,z_f)} \nonumber\\
 & & +\frac{2p/\d_f(z_o,z_f)}{1+[\,q\,\nu^2(z_f)\,]^p}\,,
\eea
where $\d_f(z_o,z_f)=\d_c D(z_o)/D(z_f)$, with $D(z)$ the linear perturbation growth function. In estimating the bias, it is sometimes simply assumed that the redshift of formation and observation coincide, $z_o=z_f$, in which case Eqn.~(\ref{eq:halo_bias}) can be substituted directly into Eqn.~(\ref{eq:bias}) to calculate the expected bias for a given cluster sample. However, this assumption is clearly not realistic. 

To treat the more plausible case $z_f>z_o$ requires a prescription for the probability distribution of the redshift of formation $z_f$ for clusters of mass $M$ observed at redshift $z_o$. This issue was first explored in \citep{LaceyCole1993,LaceyCole1994}, where the formation time of a halo observed at redshift $z_o$ is defined as the time when the most massive progenitor accreted a mass equal to half the final mass. They also provided relatively simple formulas to compute the formation redshift distribution based on the spherical collapse model. Here we make use of \cite{GiocoliEtal2007}, where an improved estimate of the formation time is proposed and compared to N-body simulations. We consider the rescaled probability for the redshift of formation $z_f$ of a halo of mass $M$ observed at redshift $z_o$ to be given by
\beq
p(\omega)=2\,\omega ~{\rm erfc}(\omega/\sqrt{2})\,,
\eeq
where 
\beq
\omega=\sqrt{q}\frac{\d_f(z_o,z_f)-\d_f(z_o,z_o)}
{\sqrt{\sigma^2(M/2,z_o)-\sigma^2(M,z_o)}}\,,
\eeq
and where, as above, $q=0.707$. The bias for halos of mass $M$ observed at redshift $z_o$ is then given by 
\bea\label{eq:halo_bias_zF}
b_h(M,z_o) & = & \int_{z_o}^{\infty}b_h(M,z_o,z_f)\times\nonumber\\
& & p[\omega(M,z_o,z_f)]
\frac{d\omega(M,z_o,z_f)}{dz}dz
\eea
where the function $b_h(M,z_o,z_f)$ is given by Eqn.~(\ref{eq:halo_bias}), and we assume $z_o=0.22$ as the mean redshift of observation for the cluster sample.

\clearpage
\begin{figure*}[t]
\begin{center}
\includegraphics[width=0.4\columnwidth]{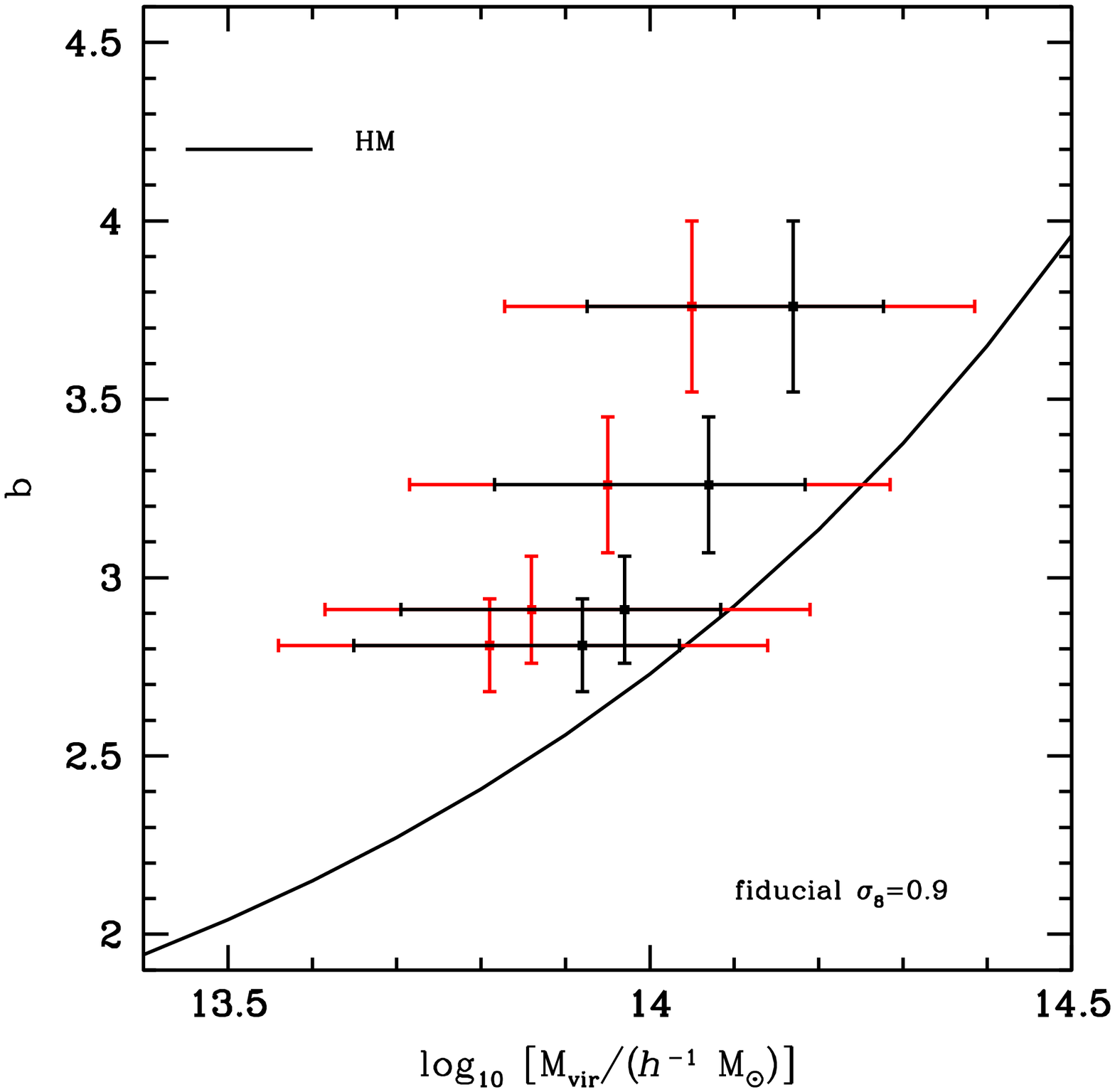}
\includegraphics[width=0.4\columnwidth]{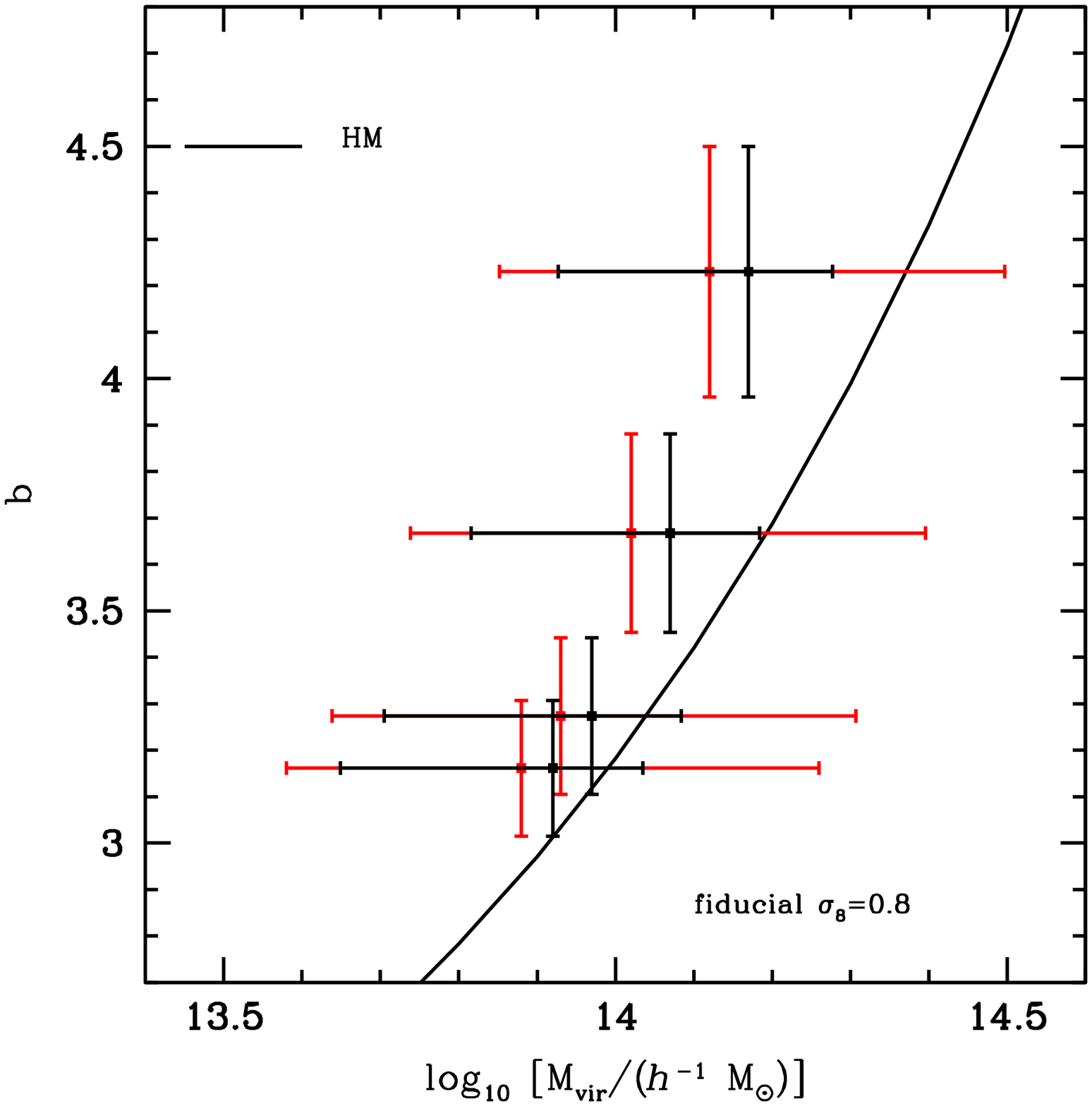}
\caption{\label{fig:biasmass} 
Bias vs. halo virial mass. Points indicate data from the cluster correlation function using the full separation range $r=20-195 \Mpc$. Cluster richness has been translated into halo virial mass using the weak lensing relation of Eqn. (\ref{eq:M200N200}). The smaller (black) horizontal error bars correspond to  the weak lensing measurement uncertainty on the constant of proportionality $M_{vir|20}$ in the mass-richness relation \citep{Johnston2007}, while the larger (red) 
horizontal error bars also include a $50\%$ scatter in halo mass at fixed richness. The 
position of the data point itself is the mean value in both cases; the 
inferred mass is slightly smaller when the intrinsic scatter is taken into account. Solid curve shows halo 
bias prediction, $b_h(M,z)$, from the Halo Model, using the mean cluster redshift $z=0.22$. {\it Left panel:} assuming $\sigma_8=0.9$; {\it Right panel}: $\sigma_8=0.8$.}
\end{center}
\end{figure*}
\clearpage

Fig.~\ref{fig:bias} shows 
the predicted HM values for the cluster bias as a function of richness and 
compares them with the values of the bias parameters obtained from the correlation function analysis (Table \ref{tab:results_bias}). The dashed curve corresponds to the approximation $z_o=z_f$, while the continuous curve uses the more realistic treatment of Eqn.~(\ref{eq:halo_bias_zF}). The shaded areas around the curves represent the 
errors on these predictions assuming 
an uncertainty in the survey volume of $20\%$.
The right panel of Fig.~\ref{fig:bias} shows the HM results with a lower 
power spectrum amplitude than our fiducial value, $\sigma_8=0.8$. The HM prediction, including the treatment of the formation redshift, appears to be in satisfactory agreement with the cluster correlation function measurements, especially for the lower value of $\sigma_8$. We also note that our result for the $\Ng\ge 10$ sample is in good agreement with the value for the bias parameter obtained from the power spectrum analysis of the same sample in \cite{Huetsi2007}, that is, $b=3(\sigma_8/0.85)$.

In Fig.~\ref{fig:bias}, the HM prediction for halo bias vs. mass has been translated to bias vs. richness by using the number density matching condition in 
Eqn. (\ref{eq:density}). As noted above, the halo mass for given richness derived from this procedure is a factor of two higher than that inferred more directly from statistical weak lensing measurements \citep{Johnston2007}, which suggests that the HM curves in Fig.~\ref{fig:bias} should perhaps be shifted horizontally to the right by about a factor of two in $N_{200}$ (since halo mass is close to linear in richness). A plausible cause of this mismatch is the neglect of the large scatter ($\sim 50$\%) in the mass-richness relation \citep{Rozo:2007yt} when matching the number density of clusters to the theoretical density of massive halos. 
 
To address this problem, we can dispense with the number-density matching and instead make use of the weak lensing results to translate optical cluster richness to halo virial mass, i.e., we can translate the cluster correlation function measurements into a measure of bias vs. halo mass, with a minimum of theoretical assumptions. Those results can then be compared directly with the HM predictions for $b_h(M)$. Using statistical weak lensing measurements, \cite{Johnston2007} found the mean relation between halo virial mass and cluster richness to be  
\beq\label{eq:M200N200}
M_{vir}(N_{200})=M_{vir|20}\left(\frac{N_{200}}{20}\right)^{\alpha_N}\,,
\eeq
where $M_{vir|20}=1.1\times 10^{14}\Ms$ and $\alpha_N=1.29$. The combined statistical plus systematic error on $M_{vir|20}$ is about 17\%; the corresponding error on the exponent $\alpha_N$ is about 3\%. However, as noted above, the intrinsic scatter in the mass at fixed richness is larger, of order 50\%. Including that scatter, we can calculate the distribution of halo masses, $P(M_{vir})$, for each of the four cluster richness samples. For each sample, we use the mean of the $P(M_{vir})$ distribution as the effective mean halo mass and the 68\% confidence interval to denote the spread in mass.  In recent work \cite{Rozo2008} have shown that the mass calibration in 
Eqn.(\ref{eq:M200N200}) has a bias that can be corrected by boosting the virial masses by 18\%, this correction
is applied in the results presented here.

\begin{figure}[t]
\begin{center}
\includegraphics[width=0.98\columnwidth]{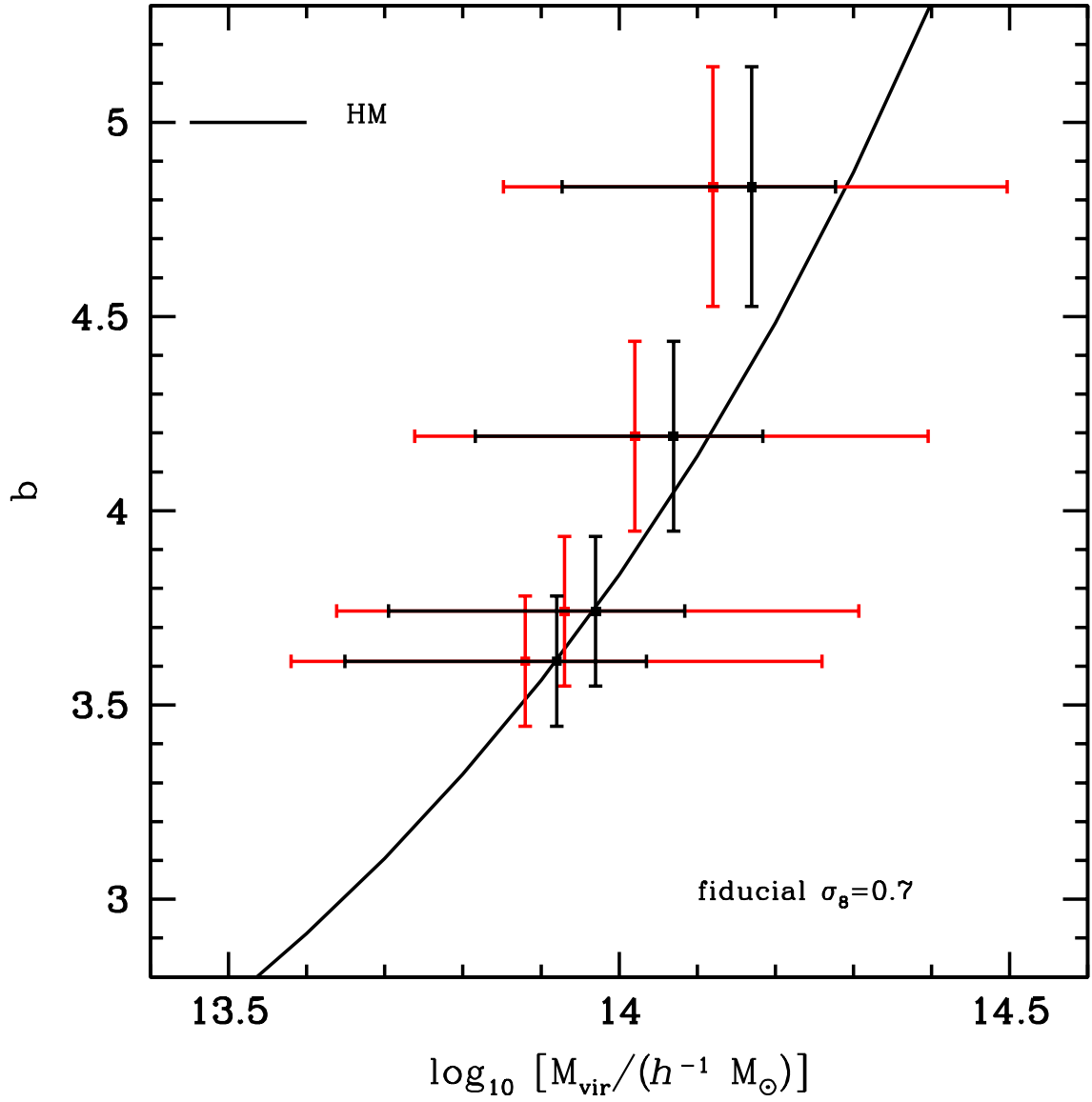}
\caption{\label{fig:biasmass7} Same as Fig. ~\ref{fig:biasmass} but for $\sigma_8=0.7$.}
\end{center}
\end{figure}

The resulting bias vs. mass relation inferred from the cluster correlation function and weak lensing mass-richness calibration is shown by the data points in Fig.~\ref{fig:biasmass} for $\sigma_8=0.8$ and $0.9$. For comparison, the linear halo bias vs. mass relation from HM is shown by the solid curve. The measured cluster bias at the central effective halo mass for each richness sample appears to be about $\sim 15-20$\% higher than the HM prediction, with a slightly lower discrepancy for lower $\sigma_8$. This is consistent with the results of \S \ref{subsec:estimate}. To further illustrate this trend, in Fig.~\ref{fig:biasmass7} we show the same comparison for $\sigma_8=0.7$, where the agreement between the measurements and the HM is further improved. The fact that the measured bias is slightly higher than that predicted from the HM could be an indicator of halo
assembly bias \citep{wu}.

\clearpage
\clearpage

\section{Conclusions}
\label{sec:conclusions}

We have measured the large-scale 3D correlation function for four richness samples in the MaxBCG catalog of optically selected galaxy clusters from the SDSS, currently the largest cluster catalog available. Since the cluster redshifts in this sample were estimated photometrically, our modeling of the observed correlation function includes a careful treatment of the impact of photo-$z$ errors. The geometric approach to photo-$z$ errors we have introduced should be of broad utility in the analysis of future large photometric galaxy surveys. 

On scales $r=20-60 \Mpc$, the cluster correlation function is well fit by a power law in separation, with a correlation scale $R_0$ that increases with cluster richness. We have determined the relation between the correlation scale $R_0$ and the mean cluster separation $d$, finding qualitative agreement with the compilation of previous cluster measurements presented in \cite{bahcall_data}. The scaling of $R_0$ with $d$ is also consistent with that predicted in N-body simulations of $\Lambda$CDM \citep{bahcall_sim}, but with a slightly ($10-15$\%) higher value of $R_0$ at fixed $d$.  

We have modeled the large-scale correlation function on scales $r=20-195 \Mpc$ 
using a non-linear model of the $\Lambda$CDM power spectrum that includes the effects of non-linear damping of the baryon acoustic peak. Non-linear damping, coupled with the estimated photo-$z$ errors, imply that we do not expect a robust detection of the BAO feature in these samples. Indeed, we find that the data set does not yield a clear detection of baryon acoustic features in the correlation function: for the largest sample considered, $\Ng\ge 10$, the significance for the best-fit BAO model with respect to a featureless model is about $1.4-1.7\sigma$, depending on whether the covariance matrix is determined using the jackknife procedure or linear perturbation theory. 

Comparison of the clustering on scales less than and greater than $r \sim 60 \Mpc$ provides weak evidence that the clustering amplitude for the $\Ng\ge 10$ and 11 samples is larger on large scales than on small scales, relative to a $\Lambda$CDM power spectrum. Such a suggestion of extra large-scale power has also been seen in the clustering of the SDSS photometric LRG sample \citep{Blake2007,Padmanabhan2007}, but statistical confirmation will require samples covering larger volumes. 

Finally, we have combined the clustering measurements with weak lensing calibration of the mass-richness relation \citep{Johnston2007} to directly infer the bias as a function of halo mass, a fundamental quantity in studies of structure formation. Again, the trend of increasing bias with halo mass is qualitatively consistent with the predictions of $\Lambda$CDM, but the amplitude of the bias is $\sim 15-20$\% higher than the model prediction. This disagreement is reduced for lower values of the power spectrum amplitude $\sigma_8$. Given the large intrinsic scatter in the relation between halo mass and cluster optical richness, however, as well as uncertainty in the photo-$z$ error dispersion $\sigma_z$, 
it is not yet clear whether this is a significant discrepancy. Moreover, 
an elevated cluster bias could be a sign of assembly bias \citep{wu}.

\acknowledgments
We express our gratitude to Martin Crocce for comments and for his help with the implementation of the RPT prescription. We also 
thank Rom\'an Scoccimarro, Jim Annis, Ben Koester, Huan Lin, Eduardo Rozo, Pasquale Serpico, Hee-Jong Seo, and Erin Sheldon for useful 
discussions and communications. We thank the referee for comments that 
significantly improved the presentation.
E.S. and J.F. acknowledge the hospitality of the Aspen Center for Physics where part of this work was completed. This research was supported by the DOE.

Funding for the creation and distribution of the SDSS and SDSS-II
has been provided by the Alfred P. Sloan Foundation,
the Participating Institutions,
the National Science Foundation,
the U.S. Department of Energy,
the National Aeronautics and Space Administration,
the Japanese Monbukagakusho,
the Max Planck Society, and the Higher Education Funding Council for England.
The SDSS Web site \hbox{is {\tt http://www.sdss.org/}.}

The SDSS is managed by the Astrophysical Research Consortium
for the Participating Institutions.  The Participating Institutions are
the American Museum of Natural History,
Astrophysical Institute Potsdam,
University of Basel,
Cambridge University,
Case Western Reserve University,
University of Chicago,
Drexel University,
Fermilab,
the Institute for Advanced Study,
the Japan Participation Group,
Johns Hopkins University,
the Joint Institute for Nuclear Astrophysics,
the Kavli Institute for Particle Astrophysics and Cosmology,
the Korean Scientist Group,
the Chinese Academy of Sciences (LAMOST),
Los Alamos National Laboratory,
the Max-Planck-Institute for Astronomy (MPA),
the Max-Planck-Institute for Astrophysics (MPiA), 
New Mexico State University, 
Ohio State University,
University of Pittsburgh,
University of Portsmouth,
Princeton University,
the United States Naval Observatory,
and the University of Washington.



\end{document}